\documentclass[12pt]{article}
\usepackage{a4wide,cite,amsmath}
\usepackage{array}
\numberwithin{equation}{section}
 \begin{document}

\def\be{\begin{equation}}
\def\ee{\end{equation}}

\def \gl{\lambda}
\def\gbar{\overline g}
\def \gs {\sigma}
\def \x{\times}
\def\vE{\overrightarrow{E}}
\def\vr{\overrightarrow{r}}
\def\vi{\bf{i}}
\def\vj{\bf{j}}
\def\vk{\bf{k}}
\def\nn{\nonumber\\}
\def\imp{\Rightarrow}
\def\pa{\partial}
\def\u{{\bf u}}
\def\r{{\bf r}}
\def\v{{\bf v}}
\def\S{{\bf S}}
\def\n{{\bf n}}
\def\tf{\tilde f}
\def\tg{\tilde g}
\def\fbar{\overline f}
\def\gbar{\overline g}
\def\pa{\partial}
\def\phibar{\overline \phi}
\def\hphi{\hat\phi}
\def\hpsi{\hat\psi}
\def\yhat{\hat y}
\def\Int{\int_{-\infty}^{\infty}}
\def\Acal{\cal{A}}
\def\go{\rightarrow}
\newcommand{\equal}{\:=\:}
%Material for preparing the first page, to include examiner's name, contact number and rubric.
%This will be incorporated by Lisbeth on the front page provided by the Central Administration.

\begin{titlepage}
\begin{flushright}
LTH1413
\end{flushright}
\date{}
\vspace*{3mm}

\begin{center}
{\Huge Asymptotic Pad\'e Predictions up to Six Loops in QCD and Eight Loops in $\lambda\phi^4$}\\[12mm]
{\bf J.A.~Gracey, I.~Jack and D.R.T.~Jones}\\
%\end{center}

\vspace{5mm}
Dept. of Mathematical Sciences,
University of Liverpool, Liverpool L69 3BX, UK\\

\end{center}

\vspace{3mm}

\begin{abstract}
 We assess the accuracy of our previous Asymptotic Pad\'e predictions of the five-loop QCD $\beta$-function and quark mass anomalous dimension in the light of subsequent exact results. We find the low-order coefficients in an expansion in powers of $N_F$ (the number of flavours) were correct to within $1\%$.
Furthermore an examination of recent results in $\lambda\phi^4$ theory indicates that the Asymptotic Pad\'e methods deliver predictions which increase in accuracy with loop order. Encouraged by this, we present six-loop Asymptotic Pad\'e predictions for the QCD $\beta$-function and quark mass anomalous dimension, and also for the eight-loop $\beta$-function in $O(N)$ $\lambda\phi^4$ theory.
\end{abstract}

\vfill

\end{titlepage}
%Change the text as appropriate.
\section{Introduction}
\label{intro1}
The method of Pad\'e approximation has been successfully used in many areas of physics such as condensed matter and statistical mechanics\cite{sam2,sam4}. Applications in quantum field theory, in particular QCD, have been accompanied by analyses of convergence properties and renormalisation-scale invariance\cite{yang,sam6,sam7,sam8,sam9,kast,drum,sam10,sam11}. Based on this analysis of convergence, a new method of Asymptotic Pad\'e Approximant Predictions (APAP) was introduced in Ref.~\cite{sam1}, which corrected the standard Pad\'e Approximant Prediction (PAP) using an asymptotic error formula. This method was applied to QCD in Ref.~\cite{chish2}, to $\lambda\phi^4$ theory (up to six-loop order) in Ref.~\cite{chish}, to supersymmetric QCD in Ref.~\cite{sqcd4} and in particular was further improved in Ref.~\cite{wap1} in the form of the Weighted Asymptotic Pad\'e Approximant Prediction (WAPAP). This new method was applied in Ref.~\cite{wap1} to predicting the then-unknown five-loop QCD $\beta$-function and quark mass anomalous dimension. Interest in Pad\'e approximation methods applied to QCD and indeed QFT has somewhat abated since then but there is still some activity, see Refs.~\cite{anant,chish3} and references within. Other significant exceptions from our point of view are the comparison in Ref.~\cite{panz} of their exact computation of the six-loop $\beta$-function in scalar $\lambda\phi^4$ theory with the prediction in Ref.~\cite{chish}, and the comparisons in Refs.~\cite{chet1,chet2} of their exact computations of the five-loop QCD $\beta$-function and quark mass anomalous dimension with the predictions in Ref.~\cite{wap1}. The authors of Refs.~\cite{panz,chet1,chet2} found that the predictions showed ``very small deviations from'', ``strikingly good agreement with'' and ``impressively good agreement with'' their new exact results; nevertheless, we shall argue that in the QCD case the agreement both for the $\beta$-function and the quark mass anomalous dimension is even better than appeared at first. This is because, as we commented in Ref.~\cite{wap1}, there were indications even at four loops that the WAPAP procedure was delivering an estimate which did not include new quartic Casimirs (QC)\cite{qcd4}. The authors of Refs.~\cite{chet1,chet2} quite naturally included the quartic Casimirs in their comparisons, but if these are omitted, the accuracy of the predictions increases dramatically (we shall give details later). Motivated by this, and also by the idea of extending the QCD and $\lambda\phi^4$ APAP and WAPAP predictions to a further loop order, we perform a thorough comparison of all the predictions in Ref.~\cite{wap1} with the now available exact results in Refs.~\cite{chet1,york1,franz1,york2} (in the case of the QCD $\beta$-function) and Refs.~\cite{chet2,york3,chet3,larin} (in the case of the quark mass anomalous dimension); we compare the accuracies of different variants of the Pad\'e procedure both for QCD and for $\lambda\phi^4$ theory, and then apply our experience to the prediction of the six-loop QCD $\beta$-function and quark mass anomalous dimension, and the eight-loop $\beta$-function in $\lambda\phi^4$ theory.

The plan of the paper is as follows: in Section~\ref{intro} we give a brief introduction to the Pad\'e approximation method and the Asymptotic Pad\'e Approximant Prediction (APAP), then in Section~\ref{qcd} we explain the Averaged Asymptotic Pad\'e Approximant Prediction (AAPAP) and explain its application to the  prediction of the (of course now known) four-loop QCD $\beta$-function. In Section~\ref{wapap} we describe a further refinement, namely the Weighted Asymptotic Pad\'e Approximant Prediction (WAPAP) method, and apply it (together with APAP and AAPAP) to the four-loop QCD $\beta$-function; and in Section~\ref{qcd5} we describe the application in Ref.~\cite{wap1} of AAPAP and WAPAP to the prediction of the five-loop QCD $\beta$-function, and use the now-available exact results to analyse the relative accuracies of APAP, AAPAP and WAPAP. In Section~\ref{qcdanom} we perform a similar exercise for the four and five-loop quark mass anomalous dimensions. In Sections~\ref{6qcd}, \ref{qcdanom6} we apply the experience acquired in previous Sections to obtaining the best possible predictions for the six-loop QCD $\beta$-function and quark mass anomalous dimension, respectively. Then in Section~\ref{6phi} we turn our attention to scalar $O(N)$ $\lambda\phi^4$ theory and review the application of Pad\'e approximants in this context up to seven loops. In Section~\ref{8phi} we present our Pad\'e predictions for the eight-loop $\beta$-function in scalar $O(N)$ $\lambda\phi^4$ theory. Finally in Section~\ref{conc} we offer some concluding remarks.
\section{Formalism}
\label{intro}
For a generic perturbative series
\be
S(x)=\sum_{n=0}^{N_{\rm max}}S_nx^n
\ee
the Pad\'e approximant $[N/M](x)$ (where we choose $N+M=N_{\rm max}$) is given by
\be
[N/M]=\frac{a_0+a_1x+\ldots a_Nx^N}{b_0+b_1(x)+\ldots b_Mx^M},
\ee
with $b_0=1$, and the other coefficients chosen so that 
\be
[N/M]=S+O(x^{N+M+1}) 
\label{padea}
\ee
(the $N$ here should not be confused with $N$ in Sections~\ref{6phi} and \ref{8phi}, where it labels the number of scalar fields).
For clarity we remark that $S_n$ will correspond to $(n+1)$-loop perturbation theory. The coefficient of the $x^{N+M+1}$ term in Eq.~\eqref{padea} is the Pad\'e Approximant Prediction (PAP) $S^{\rm{PAP}}_{N+M+1}$ of $S_{N+M+1}$. If the perturbative coefficients $S_n$ diverge as $n!$ for large $n$, it can be shown\cite{sam8} that the relative error 
\be
\delta_{N+M+1}=\frac{S^{\rm{PAP}}_{N+M+1}-S_{N+M+1}}{S_{N+M+1}}
\label{delpad}
\ee
has the asymptotic form
\be
\delta_{N+M+1}=-\frac{M!\Acal^M}{L^M_{[N/M]}}
\label{delasy}
\ee
as $N\go\infty$, for fixed $M$, where
\be
L_{[N/M]}=N+M+pM+q
\label{Lasy}
\ee
and $\Acal$, $p$, $q$ are constants. In this paper we shall be considering only the case $M=1$. Four-loop predictions are obtained as follows. For $N_{\rm max}=2$, the $[1/1]$ Pad\'e leads to the naive PAP 
\be
S_3^{\rm{PAP}}=\frac{S_2^2}{S_1}.
\ee
The improved Asymptotic Pad\'e Approximant Prediction (APAP) is then derived firstly by inserting $S_2$ (assumed known) and the $[0/1]$ Pad\'e approximation $S_2^{\rm{PAP}}=\frac{S_1^2}{S_0}$ into Eq.~\eqref{delpad} (with $N=0$) to yield an expression for $\delta_2$. We then obtain $\Acal$ from Eqs.~\eqref{delasy} and \eqref{Lasy} (again with $N=0$), choosing $p+q=0$. Rearranging Eq.~\eqref{delpad} (but now with $N=1$) yields
\be
S_3^{\rm{APAP}}=\frac{S_3^{\rm{PAP}}}{1+\delta_3}
\label{corr3}
\ee
where $\delta_3$ may be obtained using Eq.~\eqref{delasy} again, but now again with $N=1$. At higher loops exact knowledge of $\delta_3$ as well as $\delta_2$ enables us to fit both $\Acal$ and $X=p+q$. 

\section{The QCD $\beta$-function}\label{qcd}
One of the early applications of APAP\cite{sam1} was to the prediction of the then-unknown four-loop QCD $\beta$-function, so we shall use this as our first example. The QCD $\beta$-function is defined as
\be
\beta(a_s)=\frac{\pa a_s}{\pa\ln\mu^2}
\ee
where $a_s=\tfrac{\alpha_s}{4\pi}$, with $\alpha_s$ the strong coupling $\alpha_s=\tfrac{g^2}{4\pi}$, and expanded as
\be
\beta=-\sum_{n=0}\beta_na_s^{n+2},
\ee
where the lowest-order coefficients are given by\cite{gross,pol,tim,cas,tara}
\begin{align}
\beta_0=&\tfrac{11}{3}C_A-\tfrac{4}{3}T_FN_F,\nn
\beta_1=&\tfrac{34}{3}C_A^2-4C_FT_FN_F-\tfrac{20}{3}C_AT_FN_F,\nn
\beta_2=&\tfrac{2857}{54}C_A^3+N_F\left[2C_F^2T_F-\tfrac{205}{9}C_FC_AT_F-\tfrac{1415}{27}C_A^2T_F\right]\nn
&+N_F^2\left[\tfrac{44}{9}C_FT_F^2+\tfrac{158}{27}C_AT_F^2\right],
\end{align}
with $N_C$ the number of colours and $N_F$ the number of flavours (all results for renormalisation-group quantities, here and elsewhere, are computed in the usual $\overline{\rm MS}$ scheme). Of course for QCD $N_C=3$ is the value of physical interest, but we shall often list results for a range of values of $N_C$, in order to explore the behaviour of the Pad\'e approximation.
The quadratic Casimirs are defined for the case $SU(N_C)$ by
\be
C_A=N_C,\quad C_F=\frac{N_C^2-1}{2N_C},
\label{cas}
\ee
and with standard normalisation we have $T_F=\tfrac12$. (Note that for ease of comparison we have chosen our conventions to match those of Refs.~\cite{wap1,sam1} which in turn follow those of Ref.~\cite{qcd4}.)
In Ref.~\cite{sam1} an APAP prediction was made for the four-loop QCD $\beta$-function $\beta_3$, before the exact result had been derived; the results are shown in their Table II. 
The coefficient $\beta_3$ was written in the form 
\be
\beta_3=\beta_{3,0}+\beta_{3,1}N_F+\beta_{3,2}N_F^2+\beta_{3,3}N_F^3
\label{poly}
\ee
and then the APAP results for integer values $0\le N_F\le 4$  were fitted to a polynomial of this form, using a value for $\beta_{3,3}$ known already from large-$N_F$ calculations\footnote{Of course to fit the three coefficients $\beta_{3,0}$, $\beta_{3,1}$, $\beta_{3,2}$ we need at least three values of $N_F$ in the fit, but it is not clear what is the optimum number; one advantage of the WAPAP procedure described later is that it provides a criterion for this. One may also choose not to input $\beta_{3,3}$ and solve for it instead, in which case one needs four values of $N_F$ in the fit. For comparison purposes we have followed Ref.~\cite{sam1} here, but in later Sections we shall adopt a more minimal choice of range, as we explain in Section~\ref{qcd5}.}. There is a slight subtlety here: a single value for $\Acal$ (in Eq.~\eqref{delasy}) and thereby $\delta_3$ was obtained by averaging over the values of $\Acal$ obtained for each value of $N_F$ in the fitting range, and then this single value of $\delta_3$ was used to compute $S_3^{\rm{APAP}}$ (as given by Eq.~\eqref{corr3}) for each value of $N_F$; since this procedure was claimed\cite{sam1} to give better results in $\lambda\phi^4$ theory (we shall review this reasoning in Section~\ref{6phi}). We shall refer to this as the ``Averaged Asymptotic Pad\'e Approximant Prediction'', or AAPAP. The name APAP will henceforth be reserved for the alternative procedure of computing $\delta_3$ (and higher-order $\delta$s at higher loops) for each value of $N_F$ using the value of $\Acal$ appropriate for that particular value of $N_F$. We now have the capacity to make a detailed comparison of AAPAP and APAP at four and indeed five loops; we shall return to this later. We simply note here that our own four-loop calculations agree precisely with the APAP and AAPAP predictions given in Table II of Ref.~\cite{sam1}; the errors in our predictions are tabulated later (for $N_C=3$) in the fourth column of Table~\ref{qcd4} and the third column of Table~\ref{qcd4a}. In Table I of Ref.~\cite{wap1} the four-loop AAPAP predictions for the QCD $\beta$-function from  Ref.~\cite{sam1} (there labelled as ``APAP'') were compared with the exact results subsequently published in Ref.~\cite{qcd4}. For reference purposes we give these computed results for $\beta_{3,0}$, $\beta_{3,1}$, $\beta_{3,2}$, $\beta_{3,3}$ (as defined in Eq.~\eqref{poly}) here:
\begin{align}
\beta_{3,0}=&C_A^4\left(\tfrac{150653}{486}-\tfrac{44}{9}\zeta_3\right)+\tfrac{d_{AA}}{N_A}\left(-\tfrac{80}{9}+\tfrac{704}{3}\zeta_3\right),\nn
\beta_{3,1}=&\left[C_A^3\left(-\tfrac{39143}{81}+\tfrac{136}{3}\zeta_3\right)+C_A^2C_F\left(\tfrac{7073}{243}-\tfrac{656}{9}\zeta_3\right)
+C_AC_F^2\left(-\tfrac{4204}{27}+\tfrac{352}{9}\zeta_3\right)+46C_F^3\right]T_F\nn
&+\tfrac{d_{AF}}{N_A}\left(\tfrac{512}{9}-\tfrac{1664}{3}\zeta_3\right),\nn
\beta_{3,2}=&\left[C_A^2\left(\tfrac{7930}{81}+\tfrac{224}{9}\zeta_3\right)+C_F^2\left(\tfrac{1352}{27}-\tfrac{704}{9}\zeta_3\right)
+C_AC_F\left(\tfrac{17152}{243}+\tfrac{448}{9}\zeta_3\right)\right]T_F^2\nn
&+\tfrac{d_{FF}}{N_A}\left(-\tfrac{704}{9}+\tfrac{512}{3}\zeta_3\right),\nn
\beta_{3,3}=&\left[\tfrac{424}{243}C_A+\tfrac{1232}{243}C_F\right]T_F^3,
\label{bet3vals}
\end{align}
where
\be
d_{AA}=d^{abcd}_Ad^{abcd}_A,\quad d_{AF}=d^{abcd}_Ad^{abcd}_F,\quad d_{FF}=d^{abcd}_Fd^{abcd}_F.
\label{quarts}
\ee
Here $\zeta_n$ is the Riemann $\zeta$-function, and $N_A$ is the number of group generators, so that for $SU(N_C)$ we have $N_A=N_C^2-1$.    The quantities $d_{AA}$ etc as defined in Eq.~\eqref{quarts} are the quartic Casimirs which appear for the first time at four loops. They are defined in terms of the  symmetric tensors $d^{abcd}_A$, $d^{abcd}_F$ which in turn are defined in general in Ref.~\cite{qcd4}, where explicit expressions for $SU(N_C)$ are also given. The AAPAP predictions were found to be surprisingly accurate and in most cases better than might have been expected\footnote{Note that the AAPAP accuracies in Table I of Ref.~\cite{wap1} are computed using the rounded numbers given in that Table (and also contain a sign error in the second row), and therefore differ slightly from the third column of the current Table~\ref{qcd4a}, where the same accuracies are computed using unrounded results.}; however this good agreement relied on neglecting the contributions to the exact $\beta$-function from 
these new quartic Casimirs. As was argued in Ref.~\cite{wap1}, since these group structures are absent from the lower-order input data used in the Pad\'e approximation, it is not surprising that the Pad\'e procedure fails to predict these terms. It is natural to inquire into the size of these quartic Casimir contributions relative to the other parts of the $\beta$-function coefficients; in fact $d_{AA}$ contributes only to $\beta_{3,0}$, $d_{AF}$ contributes only to  $\beta_{3,1}$, and $d_{FF}$ contributes only to $\beta_{3,2}$, at around $20\%$, $10\%$ and $2\%$ respectively for $N_C=3$, with a modest decrease in the $d_{AA}$ and $d_{AF}$ contributions and increase in the $d_{FF}$ contribution with increasing $N_C$. These contributions are perhaps larger than one might have expected but, as we see, decrease with the power of $N_F$ (i.e. from $\beta_{3,0}$ through $\beta_{3,1}$ and $\beta_{3,2}$); both features which we will see again at the next loop order. Terms with $\zeta_3$ also appear for the first time at four loops but interestingly the good agreement relied on retaining these (except of course when also multiplying quartic Casimirs). This effect is even more pronounced when one considers scalar $\lambda\phi^4$ theory. The AAPAP method was also applied in Ref.~\cite{sqcd4} to the prediction of a single coefficient left undetermined in the earlier computation of the four-loop gauge $\beta$-function in supersymmetric QCD\cite{sqcd4a} (in fact, as observed in Ref.~\cite{sqcd4}, the prediction was not very sensitive to whether AAPAP or APAP was used). When subsequently calculated in Ref.~\cite{sqcd4b}, the exact value was within $10\%$ of this prediction. Surprisingly (in view of our recent remarks about $\zeta_3$), this agreement was closer when $\zeta_3$ terms in the exact four-loop result were neglected.

\section{Weighted Asymptotic Pad\'e Approximant Prediction}\label{wapap}
A variation on the APAP/AAPAP procedure was introduced in Ref.~\cite{wap1}. Firstly it was observed that the signs of $\beta_{3,0}$, $\beta_{3,1}$ and $\beta_{3,2}$ in Eq.~\eqref{bet3vals} alternate (with $T_F=\tfrac12$ and for Casimir values in Eq.~\eqref{cas} corresponding to a wide range of $N_C$, at least $N_C\le20$), implying that the APAP/AAPAP predictions for $N_F\sim 5$ are sensitive to cancellations. On the other hand, the numerical analysis is relatively stable for $N_F<0$. It turns out that more accurate predictions for $\beta_{3,0}$, $\beta_{3,1}$ and $\beta_{3,2}$ are typically obtained if one fits over a range of negative values of $N_F$ (though the accuracy of $<1\%$ for $\beta_{3,1}$ visible in Table I of Ref.~\cite{wap1}, and in the third column of our Table~\ref{qcd4a}, for AAPAP applied to a positive range of values of $N_F$, is only surpassed when one combines a negative range with the more sophisticated procedure about to be described). 
The question then is what range of negative values to use. The following procedure was found in Ref.~\cite{wap1} to give good results. 

We choose a range $-N_F^{\rm max}\le N_F\le0$ over which we determine values of $\Acal$ for each value of $N_F$  using the APAP formulae, and (as in AAPAP) we determine the arithmetic mean of these values of $\Acal$. We then use this mean $\Acal$ to estimate $\beta_3$ for each value of $N_F$ in the range
$-N_F^{\rm max}\le N_F\le0$, and do a fit to the polynomial form Eq.~\eqref{poly}\footnote{We have experimented with using the simpler APAP alternative of using each time the value of $\Acal$ pertaining to the current value of $N_F$, but found no evidence of any improvement at four or five loops.}. It is assumed that the most accurate results for $\beta_{3,0}$, $\beta_{3,1}$ and $\beta_{3,2}$ are obtained when they contribute with equal weights to the fit. For a given $N_F^{\rm max}$, these weights are $\beta_{3,0}$, $\beta_{3,1}N_F^{\rm max}/2$ and 
$\beta_{3,2}N_F^{\rm max}(2N_F^{\rm max}+1)/6$. We then estimate $\beta_{3,1}$ as follows: We take the two values of $\beta_{3,1}$ corresponding to the values of $N_F^{\rm max}$ for which the $\beta_{3,0}$ and $\beta_{3,1}$ weights are most nearly equal. We call these values of $\beta_{3,1}$, $\beta_{3,1}^{(1)}$ and $\beta_{3,1}^{(2)}$, and the corresponding weights $\beta_{3,1}^{W(1)}$ and $\beta_{3,1}^{W(2)}$. Our prediction for $\beta_{3,1}$ is then
\be
\beta_{3,1}=\frac{\Delta_2\beta_{3,1}^{(1)}+\Delta_1\beta_{3,1}^{(2)}}{\Delta_1+\Delta_2}
\label{Bcalc}
\ee
where  $\Delta_{1,2}=|\beta_{3,1}^{W(1,2)}-\beta_{3,0}^{W(1,2)}|$, with $\beta_{3,0}^{W(1,2)}$ defined similarly to $\beta_{3,1}^{W(1,2)}$. $\beta_{3,2}$ is estimated in similar fashion. Both the $\beta_{3,1}$ and $\beta_{3,2}$ calculations give values for $\beta_{3,0}$ in a similar fashion to Eq.~\eqref{Bcalc}, and we take as our prediction for $\beta_{3,0}$ the mean of these two values. This is the Weighted Average Pade Approximant Prediction procedure, or WAPAP.

It was shown in Ref.~\cite{wap1} that the WAPAP procedure worked extremely well at four loops when the $\beta_{3,3}$ coefficient was input. Once again, the evidence is displayed in Table I of Ref.~\cite{wap1}, and also in the seventh column of our Table~\ref{qcd4a}. However, it turns out that it does not work nearly as well when no coefficient is input; witness the sixth column of Table~\ref{qcd4a}. 

\begin{table}
\begin{center}
\begin{tabular}{|c|c||c|c|c|c|c|c|}
\hline
&Exact&APAP&APAP&APAP&APAP\\
&&no input&input $\beta_{3,3}$&no input&input $\beta_{3,3}$\\
&&$0\le N_F\le 4$&$0\le N_F\le 4$&$-4\le N_F\le 0$&$-4\le N_F\le 0$\\
\hline
$\beta_{3,0}$&$24633$&$-2.3\%$&$-2.2\%$&$-2.3\%$&$-2.2\%$\\
$\beta_{3,1}$&$-6375$&$1.6\%$&$3.1\%$&$2.2\%$&$1.5\%$\\
$\beta_{3,2}$&$398.5$&$-28\%$&$-10\%$&$-18\%$&$-9.5\%$\\
$\beta_{3,3}$&$1.499$&$770\%$&I/P&$-360\%$&I/P\\
\hline
\end{tabular}
\end{center}
\begin{center}
\caption{\label{qcd4}Exact results and accuracies of predictions for four-loop QCD $\beta$-function coefficients as defined in Eq.~\eqref{poly}, for $N_C=3$ and with fitting over given range of values of $N_F$ }
\end{center}
\end{table}

\begin{table}
\begin{tabular}{|c|c|c|c|c|c|c|}
\hline
&AAPAP&AAPAP&AAPAP&AAPAP&WAPAP&WAPAP\\
&no input&input $\beta_{3,3}$&no input&input $\beta_{3,3}$&no input&input $\beta_{3,3}$\\
&$0\le N_F\le 4$&$0\le N_F\le 4$&$-4\le N_F\le 0$&$-4\le N_F\le 0$&&\\
\hline
$\beta_{3,0}$&$-4.4\%$&$-4.4\%$&$-1.8\%$&$-1.8\%$&$0.65\%$&$0.13\%$\\
$\beta_{3,1}$&$-1.7\%$&$-0.34\%$&$1.5\%$&$1.1\%$&$2.1\%$&$0.13\%$\\
$\beta_{3,2}$&$-29\%$&$-13\%$&$-19\%$&$-14\%$&$-21\%$&$1.6\%$\\
$\beta_{3,3}$&$670\%$&I/P&$-210\%$&I/P&$-340\%$&I/P\\
\hline
\end{tabular}
\caption{\label{qcd4a} Exact results and accuracies of predictions} for four-loop QCD $\beta$-function coefficients (continued)
\end{table}

It therefore seemed sensible to reassess our reasons for using WAPAP in order to decide whether they are universally valid. 
In Tables~\ref{qcd4} and \ref{qcd4a} we have performed a systematic comparison of WAPAP with the  two simpler procedures APAP and AAPAP as defined in Section~2 (we referred to some of these results earlier, while discussing Refs.~\cite{sam1,wap1}). The numerical values of the exact results from Eq.~\eqref{Bcalc} are given in the first column of Table~\ref{qcd4}, and the accuracies of the various Pad\'e predictions in the remaining columns.  The appearance of ``I/P'' here and elsewhere indicates where a coefficient has been input as an exact value rather than estimated by the Pad\'e procedure. We have argued that it is better to use a negative range of values for $N_F$, but this is also something we have put to the test by displaying both options. Finally, we have compared the results obtained with input of $\beta_{3,3}$, with those obtained without inputting it. In fact, we also considered inputting $\beta_{3,2}$ as well; we have not displayed the results, since they do not materially affect our conclusions. In the case of APAP and AAPAP we used the uniform range $0\le N_F\le 4$ or
$-4\le N_F\le 0$, for ease of comparison with Ref.~\cite{sam1} where the same range was used. We have displayed the results for $N_C=3$, but we have performed calculations for a range of values of $N_C$. On the whole we find fairly moderate dependence on $N_C$, except that in the case of WAPAP with input of $\beta_{3,3}$, and also AAPAP with a positive range of $N_F$, we find that the accuracies of the predictions for $\beta_{3,1}$ vary quite considerably  with $N_C$ and in fact change sign. This behaviour is responsible for the surprisingly high accuracy of the prediction for $\beta_{3,1}$ noticeable in Table~\ref{qcd4a}. We shall later see several similar instances where monotonic behaviour of accuracies as we vary $N_C$ or $N_F$, and a resulting change of sign, leads to a high accuracy which in a sense is a fortunate accident.

Summarising the results over the whole range of values of $N_C$ tested, we found that in this current case of four-loop predictions it was typically more accurate to use a negative range for $N_F$, except in the case of $\beta_{3,1}$ where a positive range was often better (a consequence of the unusual behaviour of this coefficient mentioned in the previous paragraph). We also found that (especially in the case of a negative range) AAPAP is almost always somewhat better than APAP for low powers of $N_F$ ($\beta_{3,0}$ and $\beta_{3,1}$); while WAPAP is almost always considerably better than AAPAP. Inputting $\beta_{3,3}$ does tend to improve the predictions, as might be expected, and this effect is especially marked in the case of WAPAP. On balance, therefore, it seems that we were correct to fix on WAPAP as an improvement over APAP and AAPAP. Based on the evidence so far we shall henceforth use a negative fitting range in our discussion of the QCD $\beta$-function, unless otherwise stated.

\begin{table}[h]
\begin{center}
\begin{tabular}{|c|c|c|c|}
\hline
&APAP&AAPAP&WAPAP\\
\hline
No input&$-19\%$&$-16\%$&$-12\%$\\
$\beta_{3,3}$ input&$-13\%$&$-13\%$&$0.72\%$\\

\hline
\end{tabular}
\end{center}
\caption{\label{fullbet4}Accuracies of APAP, AAPAP and WAPAP predictions for full four-loop QCD $\beta$-function $\beta_3(N_F)$ for $N_C=N_F=3$.}
\end{table}

Arguably in practice what matters is the accuracy of the prediction for the entire $\beta$-function, $\beta_3$ in Eq.~\eqref{poly}, for $N_C=3$ and for $N_F\le 6$. In Refs.~\cite{sam1,wap1} graphs were displayed showing the predictions for the full four-loop $\beta$-function $\beta(N_F)$ over a range of values of $N_F$. WAPAP was seen to do well here, as it did for the individual coefficients in Eq.~\eqref{poly4}, for instance giving an accuracy of $0.7\%$ for $N_F=3$ (when inputting $\beta_{3,3}$). In Table~\ref{fullbet4} we go into more detail, comparing the accuracies of the predictions of the full four-loop QCD $\beta$-function obtained using APAP, AAPAP and WAPAP for $N_C=N_F=3$, and for different numbers of inputs (here, as in similar tables later, we pick $N_F=3$ as an example of a phenomenologically relevant value of $N_F$). We again see here that the success for WAPAP noted in Refs.~\cite{sam1,wap1} in the one-input case is not repeated for no inputs.  We also mention here that the predictions for $\beta_3(N_F)$ become far less accurate for increasing $N_F$; though the one-input WAPAP prediction remains within $5\%$ for $N_F\le5$.  Finally, we introduced the notion of fitting over a negative range of $N_F$ in order to improve our predictions for the coefficients of powers of $N_F$ in $\beta_3$ in  Eq.~\eqref{poly4}. But if we are only interested in the values of $\beta_3(N_F)$ for a limited number of (positive!) values of $N_F$, one might wonder if we would do just as well simply to use the Pad\'e estimates of $\beta_3$ for those particular values, and thereby avoid the whole fitting process. In fact, this almost always results in better predictions than the no-input procedures, but is easily beaten by the one-input versions, over the range $0\le N_F\le5$ (not surprisingly, given that of course we are feeding in more information in the one-input case). The one-input WAPAP is a clear overall winner for $0\le N_F\le 6$. However, this question merits further discussion in the case of scalar $\lambda\phi^4$ theory, when we shall find ourselves using a positive range of $N_F$. We shall therefore return to it later, in Section~\ref{6phi}.

\section{Five-loop QCD $\beta$-function}\label{qcd5}
Since we now have available the five-loop $\beta$-function\cite{chet1,york1,franz1,york2}  in QCD, we can check the accuracy of our predictions for this quantity, which we presented in Ref.~\cite{wap1}. These predictions were obtained using WAPAP. In the previous section we re-examined the evidence from the four-loop QCD $\beta$-function calculation which led us to select WAPAP as the most accurate procedure and confirmed that this was the most promising option based on the available evidence at the time, as we saw in Tables~\ref{qcd4} and {\ref{qcd4a}. Now we can investigate whether WAPAP was indeed the best method to use at five loops, by comparing with the other available procedures, namely APAP and AAPAP. In the case of APAP, AAPAP we decide to retain a uniform negative fitting range of $-N_{\rm min}\le N_F \le 0$ where $N_{\rm{min}}$ is the minimum value suitable for both input options (since  the evidence in favour of a negative range from our four-loop calculations was not completely conclusive, we did also perform a couple of five-loop tests with a positive fitting range, finding here woefully inaccurate results). In general we have $N_{\rm{min}}=L-1$ at $L$ loops, so that in the present five-loop case $N_{\rm{min}}=4$. This is a somewhat arbitrary choice and the reader might wonder whether more accuracy would be delivered by a fit over a larger range of values of $N_F$, which would increase the amount of data used. Arguably WAPAP offers the most plausible criteria for selecting a fitting range if one allows oneself a completely free choice, but remaining in the simpler APAP/AAPAP fixed-range context we have also experimented with a moderate increase in $N_{\rm min}$, say to $N_{\rm min}=6$, since this is the limit of phenomenological interest. In fact we have found no evidence of any consistent and sustained improvement in the accuracy of any given coefficient as we increase the fitting range at five loops (though in the AAPAP case, the accuracy of $\beta_{4,0}$ does seem to be improved by a moderate increase in the range to $N_F\approx10$). We have therefore decided to adopt the same policy of using a uniform but minimal fitting range when we come to making higher-loop predictions in later Sections. However we have continued to make trials of slightly larger ranges and we shall report on those in due course; and we should also say that at four loops, though a slightly increased range again improved the AAPAP results, it did not surpass the accuracy attained in WAPAP.

For the sake of completeness, we start by reviewing the extension of the Pad\'e procedure to this loop order.
At five loops the standard $[2/1]$ Pad\'e leads to the estimate 
 \be
\beta_4^{\rm{PAP}}=\frac{\beta_3^2}{\beta_2},
\label{b4pap}
\ee
which is then corrected in a similar fashion to Eq.~\eqref{corr3}
\be
\beta_4^{\rm{APAP}}=\frac{\beta_4^{\rm{PAP}}}{1+\delta_4}
\label{corr4}
\ee
where $\delta_4$ is given asymptotically according to Eqs.~\eqref{delasy}, \eqref{Lasy} by
\be
\delta_4=-\frac{\Acal}{L_{[2/1]}}=-\frac{\Acal}{3+X}.
\label{del4}
\ee
The constants $\Acal$ and $a+b$ may be deduced from the lower-order errors $\delta_2$ and $\delta_3$ which are defined in Eq.~\eqref{delpad} and which satisfy 
\be
\frac{\Acal}{\delta_2}=-(1+X),\quad \frac{\Acal}{\delta_3}=-(2+X).
\label{Adel}
\ee
The remainder of the procedure is much the same as at four loops, whether we are considering APAP, AAPAP or WAPAP. We are now fitting to a $\beta$-function of the form
\be
\beta_4=\beta_{4,0}+\beta_{4,1}N_F+\beta_{4,2}N_F^2+\beta_{4,3}N_F^3+\beta_{4,4}N_F^4.
\label{poly4}
\ee
We shall follow Ref.~\cite{wap1} in exploring the predictions obtained either by inputting the value of $\beta_{4,4}$ (which was already known via large-$N_F$ calculations\cite{jag2}) or by leaving it as an unknown. In the former scenario we need at least four values in our range of $N_F$, in the latter case at least five. For future reference, the value of $\beta_{4,4}$ is given by
\be
\beta_{4,4}=-\tfrac{4}{243}[(288\zeta_3+214)C_F+(480\zeta_3-229)C_A]T_F^4
\ee
(of course, we may now also obtain this from the complete results presented in Refs.~\cite{chet1,york1,franz1,york2}). 

We have performed the comparison with AAPAP for both input scenarios (we shall consider APAP shortly). We start by reproducing in Table~\ref{qcdex} the exact results obtained in Refs.~\cite{chet1,york1,franz1,york2}; these were extracted from the helpful ancillary files provided with Ref.~\cite{york2} (we have inserted numerical values for transcendental numbers such as $\zeta_3$, etc). Then in Table~\ref{qcd5b1} we show the percentage errors obtained without feeding in the values of any coefficients, for AAPAP and WAPAP respectively. The WAPAP results here correspond to Table~IV in Ref.~\cite{wap1}; the AAPAP scenario was not considered in Ref.~\cite{wap1}.  In Table~\ref{qcd5a1} we show results obtained feeding in the value of $\beta_{4,4}$, for AAPAP and WAPAP respectively. These WAPAP results correspond to Table~III in Ref.~\cite{wap1}. Throughout these comparisons we have omitted quartic Casimirs, both in the four-loop results used in making the WAPAP predictions and in the exact five-loop results used to compare with the WAPAP predictions (there are no new  higher-order Casimirs at five loops); so that in particular, the exact results displayed in Table~\ref{qcdex} are obtained by omitting any new quartic Casimirs. We call this the ``without Quartic Casimir'' or ``w/o~QC'' version of the procedure. However, we have checked that including the quartic Casimirs (the ``with Quartic Casimir'' or ``w.~QC''  version) results in a significant loss of accuracy in the $\beta$-function predictions, as was already found at four loops; here, $\beta_{4,0}$ is predicted especially badly, and the accuracy is rarely better than $10\%$ for any coefficient and any value of $N_C$. At this loop order it is no longer so intuitively clear why this should be so, since the quartic Casimirs are not making their first appearance at this level\footnote{As we commented in Section~\ref{intro1}, the authors of Ref.~\cite{chet1} compared their newly-computed exact results (for the case $N_C=3$) with ours, but including the quartic Casimirs throughout. The corresponding exact numbers in the $N_C=3$ column in Table~\ref{qcdex} would be $537148,-186162,17568,-231.3,-1.8425,$ which correspond to those in $\beta_4$ in their Eq.~(6), after including a normalisation factor of $4^L$ at $L$ loops. They have compared these exact results with the ``w.~QC'' predictions in Table III of Ref.~\cite{wap1} which they have recorded in their $\beta_4^{\rm{APAP}}$, again after accounting for a normalisation factor of $4^5$ (note that they are using the predictions where we input $\beta_{4,4}$). The percentage errors in the coefficients are $41\%$, $17\%$, $18\%$, $78\%$, which they describe as strikingly good agreement. We, on the other hand, have compared the exact numbers given in the $N_C=3$ column of Table~\ref{qcdex} (where the quartic Casimirs are omitted) with the ``w/o QC'' predictions in Table III of Ref.~\cite{wap1} to obtain accuracies given in the $N_C=3$ column in the WAPAP part of Table~\ref{qcd5a1}. We see that the ``w/o~QC'' accuracies in Table~\ref{qcd5a1} are hugely better than the ``w.~QC'' ones given a few lines back. However, anticipating the discussion later in this Section, the ``w/o~QC''  accuracies for the full $\beta$-function $\beta_4(N_F)$, while better than the `` w.~QC'' values given in Table I of Ref.~\cite{chet1}, are nevertheless increasingly poor for larger values of $N_F$.}. As at the previous loop order, it is interesting to examine the size of the contributions of the quartic Casimirs relative to the other parts of the $\beta$-function coefficients. We find that for $N_C=3$, $d_{AA}$ (as defined in Eq.~\eqref{quarts})  contributes to $\beta_{4,0}$ and $\beta_{4,1}$ at around the $10\%$ level, $d_{AF}$  contributes to $\beta_{4,1}$ and $\beta_{4,2}$ at around the $10\%$ level, and $d_{FF}$ contributes to $\beta_{4,2}$ and $\beta_{4,3}$ at around the $2\%$ level. Once again there is a modest decrease in the $d_{AA}$ and $d_{AF}$ contributions and increase in the $d_{FF}$ contribution as $N_C$ is increased, and once again there is a tendency for the relative size of contribution to decrease with the power of $N_F$ (presumably the relatively high contributions to $\beta_{4,0}$ and $\beta_{4,1}$ from $d_{AA}$ and $d_{AF}$ are related to the poor five-loop ``w.~QC'' predictions for these coefficients noted earlier). Looking more closely, one might detect a slight tendency for the corresponding contributions from each quartic Casimir to decrease with loop order, especially in the case of $d_{AA}$. Finally, we remark that the quartic Casimir contributions at this loop order are all functions of $\zeta_3$, $\zeta_4$, $\zeta_5$ and rationals (in other words at most weight five in terms of $\zeta$-values), as indeed are the other contributions to the five-loop QCD $\beta$-function; whereas one might have expected a predominance of higher-weight $\zeta$s, since the diagrams contributing higher-order Casimirs are likely to be primitive.

%Feeding in $\beta_{4,4}$ does not improve the accuracy much in the WAPAP case (except for $\beta_{4,0}$), though it does for AAPAP; perhaps the counterintuitive behaviour here might again be due to the fact that $\beta_{4,3}$ and $\beta_{4,4}$ have the same sign. Feeding in both $\beta_{4,3}$ and $\beta_{4,4}$ does improve the general accuracy for both WAPAP and AAPAP (though again $\beta_{4,0}$ is something of an exception), giving results which are also quite stable with variation of $n_c$.  But perhaps the most striking overall feature is that at five loops WAPAP fails to give an  increase in accuracy in the majority of cases. To be more precise, it does give a significant improvement in the case of $\beta_{4,1}$ and $\beta_{4,4}$ in Table~\ref{qcd5b}, $\beta_{4,0}$ in Table~\ref{qcd5a} and $\beta_{4,1}$ and $\beta_{4,2}$ in Table~\ref{qcd5d}. However, $\beta_{4,3}$ in Table~\ref{qcd5a} is considerably worse. 

\begin{table}[h]
\begin{center}
\begin{tabular}{|c|c|c|c|c|c|}
\hline
$N_C$&2&3&4&5&10\\
\hline
$\beta_{4,0}$&$64353$&$488683$&$205930$&$6284500$&$2.011\times10^8$\\
$\beta_{4,1}$&$-30822$&$-158128$&$-502463$&$-1229900$&$-1.97485\times10^7$\\
$\beta_{4,2}$&$4428$&$15415$&$36918$&$72442$&$5.83086\times10^5$\\
$\beta_{4,3}$&$-99.7$&$-233$&$-418$&$-658$&$-2648$\\
$\beta_{4,4}$&$-1.15$&$-1.84$&$-2.51$&$-3.17$&$-6.43$\\
\hline
\end{tabular}
\end{center}
\caption{\label{qcdex}Exact results for five-loop QCD $\beta$-function coefficients as defined in Eq.~\eqref{poly4}, for given values of $N_C$ (w/o~QC)}
\end{table}

\begin{table}[h]
\begin{center}
\begin{tabular}{|c|c|c|c|c|c|}
\hline 
AAPAP&&&&&\\
\hline
$N_C$&2&3&4&5&10\\
\hline
$\beta_{4,0}$&$0.32\%$&$0.40\%$&$0.43\%$&$0.44\%$&$0.38\%$\\
$\beta_{4,1}$&$1.1\%$&$0.94\%$&$0.94\%$&$0.94\%$&$0.87\%$\\
$\beta_{4,2}$&$-1.5\%$&$-1.5\%$&$-1.5\%$&$-1.5\%$&$-1.6\%$\\
$\beta_{4,3}$&$-20\%$&$-24\%$&$-26\%$&$-26\%$&$-27\%$\\
$\beta_{4,4}$&$-380\%$&$-440\%$&$-470\%$&$-490\%$&$-520\%$\\
\hline
\hline
WAPAP&&&&&\\
\hline
$N_C$&2&3&4&5&10\\
\hline
$\beta_{4,0}$&$0.87\%$&$0.75\%$&$0.69\%$&$0.65\%$&$0.59\%$\\
$\beta_{4,1}$&$1.0\%$&$0.82\%$&$0.78\%$&$0.76\%$&$0.75\%$\\
$\beta_{4,2}$&$-2.1\%$&$-2.4\%$&$-2.5\%$&$-2.6\%$&$-2.7\%$\\
$\beta_{4,3}$&$27\%$&$30\%$&$31\%$&$32\%$&$34\%$\\
$\beta_{4,4}$&$-40\%$&$-39\%$&$-37\%$&$-36\%$&$-34\%$\\
\hline
\end{tabular}
\end{center}
\caption{\label{qcd5b1} Accuracies of AAPAP and WAPAP predictions for five-loop QCD $\beta$-function coefficients (no input)}
\end{table}

\begin{table}[h]
\begin{center}
\begin{tabular}{|c|c|c|c|c|c|}
\hline
AAPAP&&&&&\\
\hline
$N_C$&2&3&4&5&10\\
\hline
$\beta_{4,0}$&$0.32\%$&$0.40\%$&$0.43\%$&$0.44\%$&$0.38\%$\\
$\beta_{4,1}$&$1.3\%$&$1.0\%$&$0.98\%$&$0.96\%$&$0.98\%$\\
$\beta_{4,2}$&$-3.5\%$&$-2.5\%$&$-2.1\%$&$-1.9\%$&$-1.7\%$\\
$\beta_{4,3}$&$15\%$&$3.6\%$&$-3.0\%$&$-7.4\%$&$-17\%$\\
\hline
\hline
WAPAP&&&&&\\
\hline
$N_C$&2&3&4&5&10\\
\hline
$\beta_{4,0}$&$0.34\%$&$0.20\%$&$0.13\%$&$0.094\%$&$0.015\%$\\
$\beta_{4,1}$&$1.3\%$&$1.1\%$&$1.0\%$&$1.0\%$&$1.0\%$\\
$\beta_{4,2}$&$-5.8\%$&$-6.3\%$&$-6.5\%$&$-6.6\%$&$-6.8\%$\\
$\beta_{4,3}$&$72\%$&$74\%$&$75\%$&$75\%$&$76\%$\\
\hline
\end{tabular}
\end{center}
\caption{\label{qcd5a1} Accuracies of AAPAP and WAPAP predictions for five-loop QCD $\beta$-function coefficients ($\beta_{4,4}$ input)}
\end{table}

%Also for low powers of $N_f$ the agreement increases with $N_c$, while it decreases with $N_c $ for higher powers of $N_f$. 

 %As we explained earlier, the WAPAP prediction for $\beta_{4,0}$ was obtained by taking the average of the WAPAP predictions for $\beta_{4,0}$ derived while estimating $\beta_{4,1}$, $\beta_{4,2}$, $\beta_{4,3}$ and $\beta_{4,4}$ (in the case of Table~\ref{qcd5b1}), or $\beta_{4,1}$, $\beta_{4,2}$ and $\beta_{4,3}$ (in the case of Table~\ref{qcd5a1}). However, with the benefit of the exact results we see that the best result for $\beta_{4,0}$ is that obtained while estimating only $\beta_{4,2}$. These better results are shown in Table~\ref{qcd5c}; we see that we achieve an astonishing accuracy. All the same, as we said before, we can achieve a $0.08$ accuracy across the board simply by using APAP, and it is only in the case of inputting $\beta_{4,4}$ alone where this selective use of WAPAP estimates significantly improves on this. Furthermore, there is no evidence that this kind of selective procedure would work consistently at other orders; for instance if we inspect the WAPAP four-loop calculation, we find that the estimates for $\beta_{3,0}$ from the computations for $\beta_{3,1}$ and $\beta_{3,2}$ bracket the correct result, and so the average does indeed give a better estimate.
The results are striking and quite surprising. In general we see that the constant term ($\beta_{4,0}$) and the coefficient of the term linear in $N_F$ ($\beta_{4,1}$) are given with an impressive accuracy of $1\%$ or better, with rare exceptions, but then there is typically a rapid drop in accuracy with increasing power of $N_F$ (though the results are in general quite stable with respect to increasing $N_C$). We have checked that in most cases AAPAP is considerably better than APAP, and therefore we have refrained from displaying the APAP results. But remarkably, in the case of $\beta_{4,0}$, APAP is almost always considerably better than both AAPAP and WAPAP, giving an accuracy consistently around $0.08\%$ (we note in passing that this feature of APAP does not apply to the QCD $\beta$-function in the ``w.~QC'' case). We see from Tables~\ref{qcd5b1} and \ref{qcd5a1} that the best results for $\beta_{4,1}$ and $\beta_{4,2}$ are given for both AAPAP and WAPAP for the case without input of $\beta_{4,4}$. The AAPAP results for $\beta_{4,3}$ are improved by input of $\beta_{4,4}$ while the WAPAP ones are worsened. The AAPAP results for $\beta_{4,0}$ are more or less the same with or without inputting $\beta_{4,4}$, whereas the WAPAP results for $\beta_{4,0}$ are considerably improved by inputting it. However, as we have said, the APAP result for $\beta_{4,0}$ is better than any of these. On the other hand, for the remaining coefficients, we see that AAPAP is usually the most accurate; better than WAPAP, except in the case of $\beta_{4,1}$ in the no-input case.

\begin{table}[h]
\begin{center}
\begin{tabular}{|c|c|c|c|}
\hline
&APAP&AAPAP&WAPAP\\
\hline
No input&$-16\%$&$-1.6\%$&$-3.7\%$\\
$\beta_{4,4}$ input&$-20\%$&$-4.5\%$&$-12\%$\\

\hline
\end{tabular}
\end{center}
\caption{\label{fullbet}Accuracies of various predictions for full five-loop $\beta$-function $\beta_4(N_F)$ for $N_C=N_F=3$.}
\end{table}

 In Section~\ref{wapap} we finished by examining the accuracies of the predictions for $\beta_3(N_F)$, and we shall do similarly here. At five loops, however, the picture is less clear-cut. In Table~\ref{fullbet} we compare the accuracies of the predictions of the full five-loop QCD $\beta$-function obtained using APAP, AAPAP and WAPAP for $N_C=N_F=3$ . We see that the result with just $\beta_{4,4}$ as input is especially disappointing for WAPAP, due to the anomalously low accuracy obtained for $\beta_{4,3}$ (see Table~\ref{qcd5a1}).
On the other hand the particularly low accuracy of $\beta_{4,4}$ using AAPAP with no inputs (see Table~\ref{qcd5b1}) has still resulted in a reasonable accuracy for $\beta_4(N_F=3)$, in fact the best in the Table. However, we never beat the accuracy for $\beta_3(N_F=3)$ obtained at four loops using WAPAP and inputting $\beta_{3,3}$. As mentioned in an earlier footnote, the predictions for $\beta_4(N_F)$ become far less accurate for increasing $N_F$; though the no-input AAPAP predictions remain reasonably close (within $2\%$) for $N_F\le5$. Furthermore, simply using the Pad\'e estimates for positive values of $N_F$ makes things considerably worse (except in the AAPAP no-input case), as does increasing the fitting range beyond $N_{\rm min}$ in the case of APAP or AAPAP (of course the fitting range is determined by the procedure itself in the case of WAPAP). We also revisited the four-loop case, finding there that an increased range appears to reduce the accuracy in the no-input case and to improve it in the no-input case; but without coming anywhere near the one-input WAPAP accuracy.

It is interesting that notwithstanding the arguments in favour of WAPAP presented in Ref.~\cite{wap1}, we have seen that it is by no means guaranteed to give the best results; so we are justified in continuing to explore alternatives in the form of APAP and AAPAP. The two reductions in accuracy we have we have just mentioned as caused by modifications of the APAP/AAPAP procedures are more significant in the five-loop than in the four-loop case, and tend to increase our confidence that we are doing the right thing in taking a minimal negative range for $N_F$ when we do use APAP or AAPAP.

Finally we mention here that we have explored what happens if we drop the ``new'' $\zeta$-functions $\zeta_4$ and $\zeta_5$ which appear for the first time at five loops; or indeed if we drop all $\zeta$ functions throughout. It turns out that our five-loop predictions are much closer to the exact results if we retain transcendental numbers such as $\zeta_3$, $\zeta_4$ and $\zeta_5$ wherever they occur.

\section{The quark mass anomalous dimension in QCD}\label{qcdanom}
In this Section we present a similar discussion to the previous one, but now studying the QCD quark mass anomalous dimension at five loops. However, one significant difference will be that here it is no longer so clear that omitting the new quartic Casimirs gives more accurate predictions, and so we shall present a detailed comparison for both alternatives. The quark mass anomalous dimension is defined by
\be
\gamma=\frac{d\ln m_q}{d\ln \mu^2}=-\sum_{n=0}\gamma_na^{n+1}
\ee
where $m_q$ is the quark mass and $a=\tfrac{\alpha_s}{\pi}$. Here we are following the conventions of Refs.~\cite{wap1,sam1} which in turn follow those of Ref.~\cite{larin} and differ slightly from those of Ref.~\cite{qcd4} in using $a$ instead of $a_s$, due to its more common use in applications\cite{larin}. We have continued with this disparity of conventions here, since otherwise it would be difficult to refer back to our five-loop predictions obtained in Ref.~\cite{wap1} without confusion. We start by listing  here the results for the anomalous dimension up to four loops\cite{nacht,tarr,tara2,chet3,larin}.
\begin{align}
\gamma_0=&\tfrac{3}{4}C_F,\nn
\gamma_1=&\tfrac{1}{16}\left(\tfrac{3}{2}C_F^2+\tfrac{97}{6}C_FC_A-\tfrac{10}{3}C_FT_FN_F\right),\nn
\gamma_2=&\tfrac{1}{64}\Bigl[\tfrac{129}{2}C_F^3-\tfrac{129}{4}C_F^2C_A+\tfrac{11413}{108}C_FC_A^2\nn
&+C_F^2T_FN_F\left(-46+48\zeta_3\right)+C_FC_AT_FN_F\left(-\tfrac{556}{27}-48\zeta_3\right)-\tfrac{140}{27}C_FT_F^2N_F^2\Bigr],\nn
\gamma_3=&\gamma_{3,0}+\gamma_{3,1}N_F+\gamma_{3,2}N_F^2+\gamma_{3,3}N_F^3,
\label{gam03}
\end{align}
where
\begin{align}
\gamma_{3,0}=&\tfrac{1}{256}\Bigl[C_F^4\left(\tfrac{-1261}{8}-336\zeta_3\right)+C_F^3C_A\left(\tfrac{15349}{12}+316\zeta_3\right)
+C_F^2C_A^2\left(-\tfrac{34045}{36}
-152\zeta_3+440\zeta_5\right)\nn
&+C_FC_A^3\left(\tfrac{70055}{72}+\tfrac{1418}{9}\zeta_3-440\zeta_5\right)+\frac{d_{AF}}{d_Q}(-32+240\zeta_3)\Bigr],\nn
\gamma_{3,1}=&\tfrac{1}{256}\Bigl[C_F^3T_F\left(-\tfrac{280}{3}+552\zeta_3-480\zeta_5\right)+C_F^2C_AT_F\left(-\tfrac{8819}{27}+368\zeta_3-264\zeta_4+
80\zeta_5\right)\nn
&+C_FC_A^2T_F\left(-\tfrac{65459}{162}-\tfrac{2684}{3}\zeta_3+264\zeta_4+400\zeta_5\right)+\frac{d_{FF}}{d_Q}(64-480\zeta_3)\Bigr],\nn
\gamma_{3,2}=&\tfrac{1}{256}\Bigl[C_F^2T_F^2\left(\tfrac{304}{27}-160\zeta_3+96\zeta_4\right)+C_FC_AT_F^2\left(\tfrac{1342}{81}+160\zeta_3-96\zeta_4\right)\Bigr],\nn
\gamma_{3,3}=&C_FT_F^3\tfrac{1}{256}\left(-\tfrac{664}{81}+\tfrac{128}{9}\zeta_3\right).
\label{gam03a}
\end{align}
Here $C_A$ and $C_F$ are defined in Eq.~\eqref{cas}. As earlier, $d_{AF}$, $d_{FF}$ are the quartic Casimirs defined in Eq.~\eqref{quarts} in terms of the symmetric tensors $d^{abcd}_A$, $d^{abcd}_F$, which in turn are defined in Ref.~\cite{qcd4}. Again, $d_{AF}$, $d_{FF}$ are given for $SU(N_C)$ in Refs.~\cite{qcd4}.} Also $d_Q$ is the dimension of the quark representation, so that $d_Q=N_C$ for $SU(N_C)$.  As in the case of the QCD $\beta$-function, we have examined the size of the quartic Casimir contributions relative to the remainder of each anomalous dimension coefficient. The contributions of $d_{AF}$ and $d_{FF}$ to $\gamma_{3,0}$ and $\gamma_{3,1}$  are around $3\%$ and $2\%$ respectively  for $N_C=3$, that of $d_{AF}$ decreasing and that of $d_{FF}$ increasing (as in the QCD case) as $N_C$ is increased ($d_{AA}$ does not contribute at this order). These contributions are slightly smaller than those in the case of the QCD $\beta$-function, though the absence of $d_{AA}$ results in the exact ``w/o~QC'' and ``w.~QC'' values of the coefficients being much closer than for QCD. We shall perform a similar investigation at the next loop order. We start our Pad\'e analysis for the quark mass anomalous dimension by briefly reviewing the experience of using Pad\'e approximants at four loops in this context. Here we are largely following the calculations of Ref.~\cite{wap1}, where the Pad\'e predictions were checked against the (then) recently-computed exact four-loop results of Refs.~\cite{chet3,larin}.

\begin{table}[h]
\begin{center}
\begin{tabular}{|c|c|c||c|c|c|c|}
\hline
&Exact&Exact&APAP&APAP&APAP&APAP\\
&w/o~QC&w.~QC&no input&input $\gamma_{3,3}$&no input&input $\gamma_{3,3}$\\
&&&$-4\le N_F\le 0$&$-4\le N_F\le 0$&$0\le N_F\le 4$&$0\le N_F\le 4$\\
\hline
$\gamma_{3,0}$&$96.44$&$98.94$&$-1.6$\%/$-4.1\%$&$-1.7$\%/$-4.1\%$&$-1.6$\%/$-4.1\%$&$-1.5$\%/$-4.0\%$\\
$\gamma_{3,1}$&$-18.83$&$-19.11$&$14\%/12\%$&$16\%/-14\%$&$13\%/12\%$&$17\%/15\%$\\
$\gamma_{3,2}$&$0.2762$&$0.2762$&$83\%/83\%$&$-17$\%/$-17\%$&$80\%/80\%$&$260\%/260\%$\\
$\gamma_{3,3}$&$0.005793$&$0.005793$&$790\%/790\%$&I/P&$1400\%/1400\%$&I/P\\
\hline
\end{tabular}
\end{center}
\begin{center}
\caption{\label{gam4}Exact results and accuracies of predictions for four-loop quark mass anomalous dimension coefficients, as defined in Eq.~\eqref{gam03}, with $N_C=3$ and with fitting over given range of values of $N_F$}
\end{center}
\end{table}

\begin{table}[h]
\begin{center}
\begin{tabular}{|c|c|c|c|c|c|c|}
\hline
&AAPAP&AAPAP&AAPAP&AAPAP\\
&no input&input $\gamma_{3,3}$&no input&input $\gamma_{3,3}$\\
&$-4\le N_F\le 0$&$-4\le N_F\le 0$&$0\le N_F\le 4$&$0\le N_F\le 4$\\
\hline
$\gamma_{3,0}$&$1.6\%/-0.99\%$&$1.6\%/-1.0\%$&$-9.9\%/-12\%$&$-9.8\%/-12\%$\\
$\gamma_{3,1}$&$4.7\%/3.2\%$&$6.0\%/4.5\%$&$-7.3\%/-8.7\%$&$-5.1\%/-6.4\%$\\
$\gamma_{3,2}$&$43\%/43\%$&$-19\%/-19\%$&$21\%/21\%$&$130\%/130\%$\\
$\gamma_{3,3}$&$490\%/490\%$&I/P&$850\%/850\%$&I/P\\
\hline
\end{tabular}
\end{center}
\caption{\label{gam4a}Accuracies of predictions for four-loop quark mass anomalous dimension coefficients with $N_C=3$ (continued)}
\end{table}

The WAPAP procedure was found in Ref.~\cite{wap1} not to work well at four loops for the quark mass anomalous dimension. For $\gamma_{3,2}$ the weight difference never changes sign and so the procedure fails to deliver a critical value of $N_F$; and for $\gamma_{3,1}$ the procedure delivers critical values of $N_F$ which increase sharply with $N_C$ and do not correspond to the most accurate results. We summarise the results for the remaining two procedures, APAP and AAPAP, in Tables~\ref{gam4} and \ref{gam4a} (taking here $N_C=3$). The second and third columns in Table~\ref{gam4} list the numerical values of the exact results given in Eq.~\eqref{gam03a}, for the ``w/o~QC'' and ``w.~QC'' cases respectively. We then give the accuracies of the estimated coefficients for each procedure, with either one or no input coefficients, and for a negative or positive fitting range of $N_F$. For each entry in the Table, we give the ``w/o~QC'' accuracy first, and the ``w.~QC'' accuracy second (to be clear, it is not the estimate which varies here, just the exact result with which we compare it). The AAPAP results in the third column of Table~\ref{gam4a} correspond to the AAPAP results (there labelled ``APAP'') in the third column of Table IX of Ref.~\cite{wap1}. As we have already observed, the exact ``w/o~QC'' and ``w.~QC'' values are very close, so that the corresponding errors are also similar. The cases where the ``w/o~QC'' estimate is more accurate predominate, but it is far from universal. The second observation is that the negative range only gives a clear advantage in the case of AAPAP. This is similar but slightly more emphatic behaviour to what was observed in the case of the QCD $\beta$-function. Finally we see that AAPAP is usually more accurate than APAP, again repeating the behaviour observed for the QCD $\beta$-function. All in all, AAPAP over a negative range is almost always better than other alternatives. In Table~\ref{fullgam4} we tabulate the values of $\gamma_3(N_F)$ for $N_C=N_F=3$, focussing now on a negative fitting range $-4\le N_F\le 0$. For $N_F=4,5,6$ the corresponding accuracies for $\gamma_3(N_F)$ are typically worse. The results here seem to indicate that AAPAP with no inputs offers our best choice of procedure. Furthermore, extending the fitting range almost always reduces the accuracy except in this case of AAPAP with no inputs, where there is a modest increase of acccuracy for $N_F\le5$ as we increase the range; but only as far as a range of $[-9,0]$, where there is a change of sign of the accuracy (here and elsewhere, the variable only takes integer values within the range).

\begin{table}[h]
\begin{center}
\begin{tabular}{|c|c|c|}
\hline
&APAP&AAPAP\\
\hline
No input&$-14\%$/$-17\%$&$1.6\%$/$-2.2\%$\\
$\gamma_{3,3}$ input&$-26\%$/$-29\%$&$-5.6\%$/$-9.1\%$\\

\hline
\end{tabular}
\end{center}
\caption{\label{fullgam4}Accuracies of APAP and AAPAP predictions for full four-loop quark mass anomalous dimension $\gamma_3(N_F)$ for $N_C=N_F=3$.}
\end{table}

Now we move on to consider five loops. 
We parametrise the five-loop quark mass anomalous dimension in QCD as
\be
\gamma_4=\gamma_{4,0}+\gamma_{4,1}N_F+\gamma_{4,2}N_F^2+\gamma_{4,3}N_F^3+\gamma_{4,4}N_F^4.
\label{gam5coeff}
\ee
Once again we have extracted the exact results from the helpful ancillary files provided with Ref.~\cite{york3}. These results are given in Tables~\ref{gamex} (w/o~QC) and \ref{gamexa} (w.~QC). At this point, as we have done previously at four loops, we pause to examine the size of the contributions of the quartic Casimirs relative to the remainders of the anomalous dimension coefficients. We find that for $N_C=3$, $d_{AA}$ (as defined in Eq.~\eqref{quarts})  contributes to $\gamma_{4,0}$  at around the $5\%$ level, $d_{AF}$  contributes to $\gamma_{4,0}$ and $\gamma_{4,1}$ at around the $3\%$ level, and $d_{FF}$ contributes to $\gamma_{4,1}$ and $\gamma_{4,2}$ also at around the $3\%$ level. As before, we see a modest decrease in the $d_{AA}$ and $d_{AF}$ contributions and increase in the $d_{FF}$ contribution as $N_C$ is increased. Once again, we see a slight reduction in the quartic Casimir contribution with the power of $N_F$. However, in this case we see no obvious tendency for the corresponding contributions from each quartic Casimir to decrease with loop order. Nevertheless, it is interesting that the quartic Casimir contributions are typically somewhat smaller here than they were in the case of the QCD $\beta$-function. Finally, we remark that the quartic Casimir contributions at this loop order, in common with the other contributions to $\gamma_4$, are all functions of $\zeta_3$, $\zeta_4$, $\zeta_5$, $\zeta_6$, $\zeta_7$, $(\zeta_3)^2$ and rationals (in other words at most weight seven in terms of $\zeta$-values), with the exception of the $d_{AA}$ contribution which (perhaps fortuitously) has no dependence on $\zeta_6$. 
The coefficient $\gamma_{4,4}$ was first computed using large-$N_F$ methods in Ref.~\cite{pal}, and $\gamma_{4,3}$ was first computed similarly in Refs.~\cite{derk1,derk2}; both these coefficients were of course subsequently rederived as part of the full five-loop computation in Refs.~\cite{chet2,york3}. They are given by
\begin{align}
\gamma_{4,4}=&C_FT_F^4\left(-\tfrac{65}{5184}-\tfrac{5}{324}\zeta_3+\tfrac{1}{36}\zeta_4\right),\nn
\gamma_{4,3}=&C_FT_F^3\Bigl[\left(\tfrac{4483}{41472}+\tfrac{11}{96}\zeta_3-\tfrac{5}{16}\zeta_4+\tfrac{1}{6}\zeta_5\right)C_F\nn
&+\left(\tfrac{18667}{248832}+\tfrac{671}{2592}\zeta_3+\tfrac{17}{72}\zeta_4-\tfrac{4}{9}\zeta_5\right)C_A\Bigr].
\end{align}
We shall now explore the accuracies obtained using various scenarios where we input none, one or two of $\gamma_{4,4}$ and $\gamma_{4,3}$.
First of all, as we did in the case of the QCD $\beta$-function, in view of the incomplete success of the  negative fitting range at four loops, we performed some trials of a positive range. We find that the positive range performs terribly for APAP and also in general for the ``w.~QC'' case; but the AAPAP ``w/o~QC'' results are comparable with the negative range case, and for $\gamma_{4,0}$, actually slightly better. We refrain from presenting a complete set of Tables, but when the time comes we shall present some results for $\gamma_4(N_F=3)$ for comparison. Accordingly, for the remainder of this Section our results will refer to a negative fitting range $[-N_{\rm min},0]$ unless otherwise stated; in fact, we find no evidence that any moderate increase in the fitting range improves the accuracy of the coefficients of powers of $N_F$ at either four or five loops. In Ref.~\cite{wap1}, predictions were given at five loops using AAPAP both with and without including the quartic Casimirs at four loops. Now that we are able to compare with the exact five-loop results as given for QCD in Ref.~\cite{chet2} and for a general gauge group in Ref.~\cite{york3}, we find (in contrast to the case of the $\beta$-function) that, as for the four-loop quark mass anomalous dimension, the distinction in accuracy between omitting and retaining the quartic Casimirs throughout is not clear-cut. Again, this is possibly related to the small size of the quartic Casimir contributions to the quark mass anomalous dimension, compared with the QCD $\beta$-function case. We have therefore tabulated the accuracies of the predictions for both the w/o~QC and w.~QC cases. We also attempt to extract some useful predictions from the WAPAP procedure, despite finding similar defects as were noticed at four loops. Specifically, we find that almost always the WAPAP procedure delivers very large critical values of $N_F$ for $\gamma_{4,3}$ (in excess of $75$), and that correspondingly the predictions obtained thereby for both $\gamma_{4,3}$ and $\gamma_{4,0}$ are extremely inaccurate. We therefore simply discard the predictions corresponding to these large critical values, deriving our final prediction for $\gamma_{4,0}$ (as described in Section~\ref{wapap}) from an average over those supplied by the WAPAP calculations for $\gamma_{4,1}$, $\gamma_{4,2}$ and $\gamma_{4,4}$ (in the cases where the latter was not already an input).

We start with the ``w/o~QC''case. Firstly in Table~\ref{gamex} we list the numerical values of the exact results, again derived from Ref.~\cite{york3}. 
Then in Table~\ref{anom5e} we show the results for the AAPAP and WAPAP procedures for the quark mass anomalous dimension at five loops. In both cases we are inputting none of the coefficients. As we explained, we are omitting the WAPAP predictions with critical values greater than 75, which is most of those for $\gamma_{4,3}$; so a gap appears at these points in the Table. We see that WAPAP performs slightly worse than AAPAP for $\gamma_{4,0}$ and $\gamma_{4,2}$, and slightly better for $\gamma_{4,1}$ and $\gamma_{4,4}$. 
In Table~\ref{anom5} we show the accuracies for the AAPAP and WAPAP procedures for the quark mass anomalous dimension at five loops, this time inputting $\gamma_{4,4}$. The AAPAP accuracies in the third column of Table~\ref{anom5}  correspond to the ``w/o~QC'' AAPAP results (there labelled ``APAP'') in the third column of Table X in Ref.~\cite{wap1}. We see that WAPAP does slightly better than AAPAP for both $\gamma_{4,0}$ and $\gamma_{4,1}$, whereas $\gamma_{4,2}$ is slightly worse; except in the case of $N_C=2$ where AAPAP is better for $\gamma_{4,0}$.  In Table~\ref{anom5b} we also input $\gamma_{4,3}$ as well as $\gamma_{4,4}$. Here we see that $\gamma_{4,0}$ and $\gamma_{4,1}$ are better with WAPAP, whereas $\gamma_{4,2}$ is about the same; except in the case of $N_C=2$ where AAPAP is still better for $\gamma_{4,0}$. 

We have also examined the APAP predictions and found that they behave similarly (though less pronouncedly) to those for the $\beta$-functions: the predictions are broadly speaking less accurate than the corresponding AAPAP results, except in the case of $\gamma_{4,0}$ where they are typically somewhat better in the ``w/o~QC'' scenario and (to anticipate a little) comparable in the ``w.~QC'' scenario.

\begin{table}[h]
\begin{center}
\begin{tabular}{|c|c|c|c|c|c|}
\hline
$N_C$&2&3&4&5&20\\
\hline
$\gamma_{4,0}$&$50.3$&$506$&$2334$&$7419$&$8.12\times10^{6}$\\
$\gamma_{4,1}$&$-21.3$&$-132$&$-445$&$-1118$&$-3.00\times10^5$\\
$\gamma_{4,2}$&$1.93$&$7.44$&$18.4$&$36.5$&$2404$\\
$\gamma_{4,3}$&$0.0396$&$0.108$&$0.205$&$0.329$&$5.51$\\
$\gamma_{4,4}$&$-0.000048$&$-0.000085$&$-0.00012$&$-0.00015$&$-0.00064$\\
\hline
\end{tabular}
\end{center}
\caption{\label{gamex}Exact results for five-loop quark mass anomalous dimension coefficients, as defined in Eq.~\eqref{gam5coeff}, for given values of $N_C$ (w/o~QC)}
\end{table}

\begin{table}[h]
\begin{center}
\begin{tabular}{|c|c|c|c|c|c|}
\hline
\small{AAPAP}&&&&&\\
\hline
$N_C$&2&3&4&5&20\\
\hline
$\gamma_{4,0}$&$0.39\%$&$-2.5\%$&$-2.8\%$&$-2.7\%$&$-1.9\%$\\
$\gamma_{4,1}$&$1.7\%$&$2.1\%$&$2.5\%$&$2.9\%$&$4.1\%$\\
$\gamma_{4,2}$&$-14\%$&$-13\%$&$-12\%$&$-12\%$&$-11\%$\\
$\gamma_{4,3}$&$-57\%$&$-43\%$&$-37\%$&$-35\%$&$-33\%$\\
$\gamma_{4,4}$&$-2100\%$&$-3600\%$&$-4800\%$&$-5300\%$&$-2500$\%\\
\hline
\hline
\small{WAPAP}&&&&&\\
\hline
$N_C$&2&3&4&5&20\\
\hline
$\gamma_{4,0}$&$2.3\%$&$-4.0\%$&$-0.38\%$&$-2.9\%$&$2.6\%$\\
$\gamma_{4,1}$&$1.8\%$&$2.1\%$&$2.3\%$&$2.3\%$&$2.4\%$\\
$\gamma_{4,2}$&$-15\%$&$-14\%$&$-14\%$&$-14\%$&$-14\%$\\
$\gamma_{4,3}$&$-120\%$&$$&$$&$$&$$\\
$\gamma_{4,4}$&$1300\%$&$1200\%$&$1400\%$&$1300\%$&$1500\%$\\
\hline
\end{tabular}
\end{center}
\caption{\label{anom5e} Accuracies of AAPAP and WAPAP predictions for five-loop quark mass anomalous dimension coefficients; blank entries indicate a failure of the WAPAP procedure (no inputs, w/o~QC)}
\end{table}

\begin{table}[h]
\begin{center}
\begin{tabular}{|c|c|c|c|c|c|}
\hline
\small{AAPAP}&&&&&\\
\hline
$N_C$&2&3&4&5&20\\
\hline
$\gamma_{4,0}$&$0.39\%$&$-2.5\%$&$-2.8\%$&$-2.7\%$&$-1.9\%$\\
$\gamma_{4,1}$&$1.8\%$&$2.2\%$&$2.6\%$&$2.9\%$&$4.1\%$\\
$\gamma_{4,2}$&$-15\%$&$-14\%$&$-13\%$&$-12\%$&$-11\%$\\
$\gamma_{4,3}$&$-78\%$&$-66\%$&$-60\%$&$-55\%$&$37\%$\\
\hline
\hline
\small{WAPAP}&&&&&\\
\hline
$N_C$&2&3&4&5&20\\
\hline
$\gamma_{4,0}$&$1.3\%$&$-1.6\%$&$-2.0\%$&$-2.2\%$&$-2.4\%$\\
$\gamma_{4,1}$&$1.9\%$&$2.2\%$&$2.3\%$&$2.4\%$&$2.5\%$\\
$\gamma_{4,2}$&$$&$$&$$&$$&$$\\
$\gamma_{4,3}$&$-83\%$&$-87\%$&$-88\%$&$-88\%$&$-87\%$\\
\hline
\end{tabular}
\end{center}
\caption{\label{anom5} Accuracies of AAPAP and WAPAP predictions for five-loop quark mass anomalous dimension coefficients; blank entries indicate a failure of the WAPAP procedure ($\gamma_{4,4}$ input, w/o~QC)}
\end{table}

\begin{table}[h]
\begin{center}
\begin{tabular}{|c|c|c|c|c|c|}
\hline
\small{AAPAP}&&&&&\\
\hline
$N_C$&2&3&4&5&20\\
\hline
$\gamma_{4,0}$&$0.46\%$&$-2.5\%$&$-2.8\%$&$-2.7\%$&$-1.9\%$\\
$\gamma_{4,1}$&$0.58\%$&$1.7\%$&$2.3\%$&$2.7\%$&$4.1\%$\\
$\gamma_{4,2}$&$-5.6\%$&$-7.9\%$&$-8.9\%$&$-9.5\%$&$-11\%$\\
\hline
\hline
\small{WAPAP}&&&&&\\
\hline
$N_C$&2&3&4&5&20\\
\hline
$\gamma_{4,0}$&$1.8\%$&$-1.0\%$&$-1.5\%$&$-1.6\%$&$-1.6\%$\\
$\gamma_{4,1}$&$0.090\%$&$0.24\%$&$0.34\%$&$0.37\%$&$0.35\%$\\
$\gamma_{4,2}$&$6.5\%$&$8.5\%$&$9.3\%$&$9.6\%$&$10\%$\\
\hline
\end{tabular}
\end{center}
\caption{\label{anom5b} Accuracies of AAPAP and WAPAP predictions for five-loop quark mass anomalous dimension coefficients ($\gamma_{4,3}$ and $\gamma_{4,4}$ input, w/o~QC)}
\end{table}

We now turn to the ``w.~QC'' case. In Table~\ref{gamexa} we show the numerical values of the exact results, once again derived from Ref.~\cite{york3}. The accuracies for the various procedures and choices of input are then shown in Tables~\ref{anom5ea}, \ref{anom5a} and \ref{anom5ba}. The AAPAP results in Table~\ref{anom5a} correspond to the ``w.~QC'' results in Table X in Ref.~\cite{wap1}\footnote{The authors of Ref.~\cite{chet2} again compared their newly-computed exact results (for the case $N_C=3$) with our ``w.~QC'' results in Ref.~\cite{wap1}. The corresponding exact numbers in the $N_C=3$ column in Table~\ref{gamexa} are exactly those in $(\gamma_m)_4$ in their Eq.~(4.5). They have compared these exact results with the ``w.~QC'' predictions in Table X of Ref.~\cite{wap1} which they have recorded in their $(\gamma_m)_4^{\rm{APAP}}$ (note that they are again using the predictions where we input $\gamma_{4,4}$). They describe the agreement, which can be observed in the third column of our Table~\ref{anom5a}, as strikingly good.}. In general we see that (as at four loops) the errors are of the same order as in the ``w/o~QC'' case, contrasting with the case of the QCD $\beta$-function where the errors were much larger in the ``w.~QC'' case. Furthermore, we observe a tendency for $\gamma_{4,1}$ to be given more accurately than in the ``w/o~QC'' case, especially in the case of no inputs and one input. $\gamma_{4,2}$ is comparable, while $\gamma_{4,0}$ is worse, especially in the AAPAP case. The comparison between AAPAP and WAPAP is much the same as in the ``w/o Q'' case except in the case of Table~\ref{anom5ba}, where now it is only for $\gamma_{4,0}$ that WAPAP is better.

\begin{table}
\begin{center}
\begin{tabular}{|c|c|c|c|c|c|}
\hline
$N_C$&2&3&4&5&20\\
\hline
$\gamma_{4,0}$&$62.5$&$560$&$2490$&$7810$&$8.35\times10^{6}$\\
$\gamma_{4,1}$&$-24.0$&$-144$&$-482$&$-1210$&$-3.24\times10^5$\\
$\gamma_{4,2}$&$1.94$&$7.48$&$18.5$&$36.8$&$2430$\\
$\gamma_{4,3}$&$0.0396$&$0.108$&$0.205$&$0.329$&$5.51$\\
$\gamma_{4,4}$&$-0.000048$&$-0.000085$&$-0.00012$&$-0.00015$&$-0.00064$\\
\hline
\end{tabular}
\end{center}
\caption{\label{gamexa}Exact results for five-loop quark mass anomalous dimension coefficients, as defined in Eq.~\eqref{gam5coeff}, for given values of $N_C$ (w.~QC)}
\end{table}

\begin{table}
\begin{center}
\begin{tabular}{|c|c|c|c|c|c|}
\hline
&\small{AAPAP}&&&&\\
\hline
$N_C$&2&3&4&5&20\\
\hline
$\gamma_{4,0}$&$-10\%$&$-5.3\%$&$-3.3\%$&$-2.3\%$&$0.26\%$\\
$\gamma_{4,1}$&$-3.1\%$&$-0.72\%$&$0.33\%$&$0.93\%$&$2.82\%$\\
$\gamma_{4,2}$&$-12\%$&$-10\%$&$-8.7\%$&$-7.9\%$&$-5.5\%$\\
$\gamma_{4,3}$&$-62\%$&$-43\%$&$-33\%$&$-24\%$&$1.7\%$\\
$\gamma_{4,4}$&$-1800\%$&$-3600\%$&$-5100\%$&$-6900\%$&$-26000\%$\\
\hline
\hline
&\small{WAPAP}&&&&\\
\hline
$N_C$&2&3&4&5&20\\
\hline
$\gamma_{4,0}$&$-8.1\%$&$-5.0\%$&$-6.1\%$&$-0.52\%$&$-1.2\%$\\
$\gamma_{4,1}$&$-3.1\%$&$-1.0\%$&$-0.28\%$&$-0.087\%$&$0.69\%$\\
$\gamma_{4,2}$&$-12\%$&$-11\%$&$-10\%$&$-9.9\%$&$-8.9\%$\\
$\gamma_{4,3}$&$-120\%$&$$&$$&$$&$$\\
$\gamma_{4,4}$&$1400\%$&$1300\%$&$1200\%$&$1400\%$&$1300\%$\\
\hline
\end{tabular}
\end{center}
\caption{\label{anom5ea} Accuracies of AAPAP and WAPAP predictions for five-loop quark mass anomalous dimension coefficients; blank entries indicate a failure of the WAPAP procedure (no inputs, w.~QC)}
\end{table}

\begin{table}
\begin{center}
\begin{tabular}{|c|c|c|c|c|c|}
\hline
\small{AAPAP}&&&&&\\
\hline
$N_C$&2&3&4&5&20\\
\hline
$\gamma_{4,0}$&$-10\%$&$-5.3\%$&$-3.3\%$&$-2.3\%$&$0.26\%$\\
$\gamma_{4,1}$&$-3.0\%$&$-0.69\%$&$0.35\%$&$0.94\%$&$2.8\%$\\
$\gamma_{4,2}$&$-12\%$&$-11\%$&$-9.4\%$&$-8.4\%$&$-5.6\%$\\
$\gamma_{4,3}$&$-79\%$&$-66\%$&$-56\%$&$-50\%$&$-22\%$\\
\hline
\hline
\small{WAPAP}&&&&&\\
\hline
$N_C$&2&3&4&5&20\\
\hline
$\gamma_{4,0}$&$-9.6\%$&$-4.7\%$&$-2.8\%$&$-1.9\%$&$-0.49\%$\\
$\gamma_{4,1}$&$-3.0\%$&$-0.98\%$&$-0.21\%$&$0.16\%$&$0.77\%$\\
$\gamma_{4,2}$&$-10\%$&$-9.2\%$&$-8.1\%$&$-7.5\%$&$-6.4\%$\\
$\gamma_{4,3}$&$$&$$&$$&$$&$$\\
\hline
\end{tabular}
\end{center}
\caption{\label{anom5a} Accuracies of AAPAP and WAPAP predictions for five-loop quark mass anomalous dimension coefficients; blank entries indicate a failure of the WAPAP procedure ($\gamma_{4,4}$ input, w.~QC)}
\end{table}

\begin{table}
\begin{center}
\begin{tabular}{|c|c|c|c|c|c|}
\hline
\small{AAPAP}&&&&&\\
\hline
$N_C$&2&3&4&5&20\\
\hline
$\gamma_{4,0}$&$-10\%$&$-5.3\%$&$-3.3\%$&$-2.3\%$&$0.26\%$\\
$\gamma_{4,1}$&$-4.1\%$&$-1.1\%$&$0.14\%$&$0.82\%$&$2.8\%$\\
$\gamma_{4,2}$&$-2.7\%$&$-5.2\%$&$-5.6\%$&$-5.7\%$&$-5.3\%$\\
\hline
\hline
\small{WAPAP}&&&&&\\
\hline
$N_C$&2&3&4&5&20\\
\hline
$\gamma_{4,0}$&$-9.2\%$&$-4.2\%$&$-2.3\%$&$-1.4\%$&$0.087\%$\\
$\gamma_{4,1}$&$-4.8\%$&$-2.8\%$&$-2.0\%$&$-1.6\%$&$-1.0\%$\\
$\gamma_{4,2}$&$10\%$&$11\%$&$12\%$&$13\%$&$14\%$\\
\hline
\end{tabular}
\end{center}
\caption{\label{anom5ba} Accuracies of AAPAP and WAPAP predictions for five-loop quark mass anomalous dimension coefficients ($\gamma_{4,3}$ and $\gamma_{4,4}$ input, w.~QC)}
\end{table}

Just as we did for the full QCD $\beta$-function in Table~\ref{fullbet}, in Table~\ref{fullgam} we compare (for the ``w/o~QC'' case) the accuracies of the predictions of the full five-loop QCD mass anomalous dimension $\gamma_4(N_F)$ obtained using APAP, AAPAP and WAPAP for $N_C=N_F=3$. The AAPAP results are uniformly poor, presumably due to the poor predictions for the higher powers of $N_F$ seen in Tables~\ref{anom5e}, \ref{anom5} and \ref{anom5b}. In the WAPAP case, since we discarded the $\gamma_{4,3}$ prediction, we have replaced it with the corresponding AAPAP prediction. The extremely small value for APAP with only $\gamma_{4,4}$ input seems to be a result of a fortuitous cancellation for the particular value $N_C=3$, whereas the reasonable APAP accuracy obtained with $\gamma_{4,3}$ and $\gamma_{4,4}$ input is sustained across different values of $N_C$. Nevertheless, for the phenomenologically interesting $N_C=3$, the no-input APAP (and with the minimal fitting range) seems the best choice for $N_F\le 6$. Finally, as an example of the reasonably accurate AAPAP ``w/o~QC'' results for a positive fitting range mentioned earlier, for a range $0\le N_F\le4$ we find an accuracy of around $12-13\%$ for $\gamma_4(N_F=3)$, irrespective of number of inputs (and $\gamma_{4,0}$ is given to an accuracy of around $1\%$).

In Table~\ref{fullgama} we perform the same exercise for the ``w.~QC'' case. Compared with the ```w/o~QC'' case, APAP is much worse, while AAPAP is very similar and WAPAP is much the same.

\begin{table}
\begin{center}
\begin{tabular}{|c|c|c|c|}
\hline
&APAP&AAPAP&WAPAP\\
\hline
No input&$14\%$&$-17\%$&$-22\%$\\
$\gamma_{4,4}$ input&$-0.040\%$&$-18\%$&$-15\%$\\
$\gamma_{4,4}$, $\gamma_{4,3}$ input&$1.9\%$&$-14\%$&$-0.25\%$\\

\hline
\end{tabular}
\end{center}
\caption{\label{fullgam}Accuracies of various predictions for full five-loop quark mass anomalous dimension $\gamma_4 (N_F)$ for $N_C=N_F=3$ (w/o~QC).}
\end{table}

\begin{table}
\begin{center}
\begin{tabular}{|c|c|c|c|}
\hline
&APAP&AAPAP&WAPAP\\
\hline
No input&$81\%$&$-17\%$&$-16\%$\\
$\gamma_{4,4}$ input&$35\%$&$-18\%$&$-15\%$\\
$\gamma_{4,4}$, $\gamma_{4,3}$ input&$19\%$&$-14\%$&$-2.0\%$\\

\hline
\end{tabular}
\end{center}
\caption{\label{fullgama}Accuracies of various predictions for full five-loop quark mass anomalous dimension $\gamma_4 (N_F)$ for $N_C=N_F=3$ (w.~QC).}
\end{table}

\section{Prediction for six-loop QCD $\beta$-function}\label{6qcd}
In this Section we present our predictions for the six-loop QCD $\beta$-function. As we have seen already, our experience at four and five loops is that 
the most accurate results for the QCD $\beta$-function are obtained by using the lower-order results, omitting the new quartic Casimirs, as inputs to predict the next order results, in which we also omit the new quartic Casimirs. We therefore expect this ``w/o~QC'' procedure to give the most accurate results at six loops too, and we shall start by giving the predictions obtained by using as input the lower order results omitting the  contributions from the new quartic Casimirs at four and five loops. However, there may some practical utility in the predictions obtained by including the new quartic invariants at four and five loops, which one might expect to capture (though perhaps less accurately) the six-loop QCD $\beta$-function including these new invariants (though arguably omitting potential ``sextic'' invariants making their first appearance at six loops). Accordingly, we shall also present these ``w.~QC''  predictions. In a similar spirit, we shall also present the APAP predictions, since we have seen that at five loops, this procedure gave the best results for $\beta_{4,0}$ in the ``w/o~QC'' case; however, after also presenting the AAPAP and WAPAP predictions, we shall give a reasoned selection of our ``best-guess'' predictions for the most accurately-determined coefficients, in both the ``w/o~QC'' and ``w.~QC'' cases.  We parametrise the six-loop QCD $\beta$-function as
\be
\beta_5=\beta_{5,0}+\beta_{5,1}N_F+\beta_{5,2}N_F^2+\beta_{5,3}N_F^3+\beta_{5,4}N_F^4+\beta_{5,5}N_F^5.
\label{poly5}
\ee
The value for $\beta_{5,5}$ has already been derived from large-$N_F$ computations\cite{jag2}:
\be
\beta_{5,5}=-\tfrac{32}{3645}[(864\zeta_4-1056\zeta_3+502)C_F+(1440\zeta_4-1264\zeta_3-453)C_A)]T_F^5.
\ee
When we apply the APAP procedure, at this order we have the relations 
\be
\delta_2=-\frac{\Acal}{1+X},\quad \delta_3=-\frac{\Acal}{2+X},\quad \delta_4=-\frac{\Acal}{3+X},
\label{deltas}
\ee
from Eq.~\eqref{Adel} together with Eq.~\eqref{del4}. Now that we know the exact five-loop result $\beta_4$, we may now also compute $\delta_4$ using Eq.~\eqref{delpad}:
\be
\delta_4=\frac{\beta_4^{\rm{PAP}}-\beta_4}{\beta_4}
\label{del4a}
\ee
where $\beta_4^{\rm{PAP}}$ is given by Eq.~\eqref{b4pap}.
$\Acal$ and $X$ are therefore overdetermined. Since the defining relation Eq.~\eqref{delasy} is an asymptotic equation, it seems sensible to use the highest-order information available and deduce $\Acal$, $X$ simply by solving the equations for $\delta_3$ and $\delta_4$. There is evidence from $\lambda\phi^4$ theory (where we have the exact six and seven-loop results) that this is reasonable: At six loops in $\lambda\phi^4$ theory, the accuracy of the prediction for the full $O(N)$  $\beta$-function $\beta_5^{\lambda}(N)$ is better for  $N\le 4$ (and almost always better for $N_F\le 6$) if we solve for $\delta_3$ and $\delta_4$ than if we solve for $\delta_2$ and $\delta_4$ or for  $\delta_2$ and $\delta_3$. The accuracy of the predictions for coefficients of individual powers of $N$ presents a more varied picture for different pairs of $\{\delta_2,\delta_3,\delta_4\}$, but coefficients of low powers of $N$ definitely seem to be best predicted when using the pair $\{\delta_3,\delta_4\}$, and the range of variation of the predictions is well within the range of variation resulting from other choices such as APAP or AAPAP. Similar behaviour is observed in $\lambda\phi^4$ theory at seven loops; here there is a $\delta_5$ satisfying relations analogous to $\delta_4$ in Eqs.~\eqref{deltas}, \eqref{del4a}. The accuracy is almost always greater for the first few coefficients of powers of $N$, and for the value of $\beta_6^{\lambda}(N)$ for $N\le 6$, if we select the pair $\{\delta_4,\delta_5\}$. We hope that this behaviour as a function of $N$ serves as a guide for that of the QCD $\beta$-function as a function of $N_F$.

The remainder of the procedure for APAP, AAPAP and WAPAP is completely analogous to the four and five-loop cases, with one minor exception:
the critical values of $N_F$ found using the WAPAP procedure for $\beta_{5,4}$ and $\beta_{5,5}$ are exceptionally high ($N_F=85$ and $N_F=120$ respectively for the case $N_C=3$); and also the values obtained for $\beta_{5,4}$ and $\beta_{5,5}$ are very unstable and sensitive to rounding errors, even when using high precision arithmetic. We saw high critical values of $N_F$ in the case of the five-loop quark mass anomalous dimension, but this instability seems to be a new feature. Accordingly, as we did in similar circumstances in Section~\ref{qcdanom}, we have omitted these variables from the WAPAP procedure, using the corresponding AAPAP values instead. The ``w/o~QC'' results for APAP, AAPAP and WAPAP (with no input) are tabulated in Tables~\ref{qcd6P}, \ref{qcd6AP} and \ref{qcd6WP}; and (with $\beta_{5,5}$ input) in Tables~\ref{qcd6Pf}, \ref{qcd6APf} and \ref{qcd6WPf}. We have once again throughout this Section used a fit over the minimal range $-N_{\rm min}\le N_F\le 0$ for both APAP and AAPAP, where at six loops $N_{\rm min}=5$; recall that in Section~\ref{qcd5} we found no evidence of consistent improvement when we increased the range. As explained at the start of this Section, here we have omitted the quartic Casimirs from the four and five loop inputs, and correspondingly we expect these predictions to be closest to the six-loop results with quartic (and possible sextic) Casimirs omitted. We see here a remarkable agreement amongst the AAPAP and WAPAP predictions for the first three coefficients, especially for small $N_C$, with the maximum difference being around $0.3\%$. The APAP coefficients are very close too, with a maximum difference of around $1.3\%$.  Of course these predictions may all be converging on the wrong answer, but since the five-loop predictions for the first three QCD $\beta$-function coefficients are typically  more accurate than those at four loops (except for WAPAP with input of $\beta_{3,3}$), it seems reasonable to expect this trend to continue; especially as similar trends will be observed through seven loops for scalar $\lambda\phi^4$ theory in Section~\ref{6phi}. Now we shall give our ``best-guess'' predictions. We shall see later in Section~\ref{6phi} that there is evidence that Pad\'e properties alternate in loop order, and therefore we should expect six-loop behaviour to resemble that at four loops. Based on our experience in Section~\ref{qcd4}, we therefore expect WAPAP with one input to give the best predictions for the six-loop QCD $\beta$-function.  Our most confident predictions for the ``w/o~QC'' case are therefore (for $N_C=3$)
\begin{align}
\beta^{\rm{w/o~QC}}_{5,0}=&1.076\times 10^7,\quad \beta^{\rm{w/o~QC}}_{5,1}=-4.182\times 10^6,\nn
\beta^{\rm{w/o~QC}}_{5,2}=&550200,\quad \beta^{\rm{w/o~QC}}_{5,3}=-20400.
\label{6res}
\end{align}
The remaining two coefficients vary so much that it seems unlikely that we can make any useful predictions for them. Even in the case of the third and fourth coefficients, the differences between our selection, and the result of a simple average, is probably beyond any accuracy we can reasonably anticipate.
\begin{table}
\begin{center}
\begin{tabular}{|c|c|c|c|c|c|}
\hline
$N_C$&2&3&4&5&10\\
\hline
$\beta_{5,0}$&$946000$&$1.078\times10^7$&$6.055\times10^7$&$2.310\times10^8$&$1.478\times10^{10}$\\
$\beta_{5,1}$&$-552600$&$-4.230\times10^6$&$-1.790\times10^7$&$-5.476\times10^7$&$-1.758\times10^9$\\
$\beta_{5,2}$&$104600$&$543400$&$1.732\times10^6$&$4.245\times10^6$&$6.822\times10^7$\\
$\beta_{5,3}$&$-4762$&$-15730$&$-37140$&$-72910$&$-596400$\\
$\beta_{5,4}$&$209.1$&$701.9$&$1406$&$2222$&$7330$\\
$\beta_{5,5}$&$5.943$&$24.87$&$44.32$&$56.60$&$5.216$\\
$\beta_5$&$119700$&$2.614\times10^6$&$2.155\times10^7$&$1.031\times10^8$&$1.011\times10^{10}$\\
\hline
\end{tabular}
\end{center}
\caption{\label{qcd6P} APAP predictions for six-loop QCD $\beta$-function coefficients, as defined in Eq.~\eqref{poly5}, for given values of $N_C$ (no inputs, w/o~QC)}
\end{table}

\begin{table}
\begin{center}
\begin{tabular}{|c|c|c|c|c|c|}
\hline
$N_C$&2&3&4&5&10\\
\hline
$\beta_{5,0}$&$944900$&$1.077\times10^7$&$6.056\times10^7$&$2.312\times10^8$&$1.480\times10^{10}$\\
$\beta_{5,1}$&$-545000$&$-4.183\times10^6$&$-1.772\times10^7$&$-5.424\times10^7$&$-1.742\times10^9$\\
$\beta_{5,2}$&$106000$&$550600$&$1.757\times10^6$&$4.309\times10^6$&$6.936\times10^7$\\
$\beta_{5,3}$&$-5847$&$-20590$&$-49500$&$-97300$&$-784600$\\
$\beta_{5,4}$&$4.9381$&$-12.00$&$-42.75$&$-85.69$&$-459.4$\\
$\beta_{5,5}$&$-0.4043$&$-2.198$&$-4.515$&$-7.088$&$-20.84$\\
$\beta_5$&$106600$&$2.619\times10^6$&$2.187\times10^7$&$1.046\times10^8$&$1.018\times10^{10}$\\
\hline
\end{tabular}
\end{center}
\caption{\label{qcd6AP} AAPAP predictions for six-loop QCD $\beta$-function coefficients (no inputs, w/o~QC)}
\end{table}

\begin{table}
\begin{center}
\begin{tabular}{|c|c|c|c|c|c|}
\hline
$N_C$&2&3&4&5&10\\
\hline
$\beta_{5,0}$&$945100$&$1.076\times10^7$&$6.044\times10^7$&$2.306\times10^8$&$1.475\times10^{10}$\\
$\beta_{5,1}$&$-545000$&$-4.182\times10^6$&$-1.770\times10^7$&$-5.415\times10^7$&$-1.738\times10^9$\\
$\beta_{5,2}$&$106000$&$549900$&$1.753\times10^6$&$4.297\times10^6$&$6.913\times10^7$\\
$\beta_{5,3}$&$-5785$&$-20220$&$-48450$&$-95070$&$-765000$\\
$\beta_{5,4}$&$4.9381$&$-12.00$&$-42.75$&$-85.69$&$-459.4$\\
$\beta_{5,5}$&$-0.4043$&$-2.198$&$-4.515$&$-7.088$&$-20.84$\\
$\beta_5$&$108500$&$2.615\times10^6$&$2.180\times10^7$&$1.042\times10^8$&$1.014\times10^{10}$\\
\hline
\end{tabular}
\end{center}
\caption{\label{qcd6WP} WAPAP predictions for six-loop QCD $\beta$-function coefficients (no inputs, w/o~QC)}
\end{table}

\begin{table}
\begin{center}
\begin{tabular}{|c|c|c|c|c|c|}
\hline
$N_C$&2&3&4&5&10\\
\hline
$\beta_{5,0}$&$946000$&$1.078\times10^7$&$6.055\times10^7$&$2.310\times10^8$&$1.478\times10^{10}$\\
$\beta_{5,1}$&$-552900$&$-4.232\times10^6$&$-1.791\times10^7$&$-5.476\times10^7$&$-1.758\times10^9$\\
$\beta_{5,2}$&$104100$&$541100$&$1.728\times10^6$&$4.239\times10^6$&$6.822\times10^7$\\
$\beta_{5,3}$&$-5075$&$-17070$&$-39530$&$-75970$&$-596700$\\
$\beta_{5,4}$&$137.3$&$394.5$&$856.2$&$1520$&$7276$\\
$\beta_5$&$98130$&$2.522\times10^6$&$2.138\times10^7$&$1.029\times10^8$&$1.011\times10^{10}$\\
\hline
\end{tabular}
\end{center}
\caption{\label{qcd6Pf} APAP predictions for  six-loop QCD $\beta$-function coefficients (inputting $\beta_{5,5}$) (w/o~QC)}
\end{table}

\begin{table}
\begin{center}
\begin{tabular}{|c|c|c|c|c|c|}
\hline
$N_C$&2&3&4&5&10\\
\hline
$\beta_{5,0}$&$944900$&$1.077\times10^7$&$6.056\times10^7$&$2.312\times10^8$&$1.480\times10^{10}$\\
$\beta_{5,1}$&$-545000$&$-4.182\times10^6$&$-1.772\times10^7$&$-5.423\times10^7$&$-1.742\times10^9$\\
$\beta_{5,2}$&$106100$&$550800$&$1.757\times10^6$&$4.310\times10^6$&$6.936\times10^7$\\
$\beta_{5,3}$&$-5815$&$-20450$&$-49230$&$-96890$&$-783400$\\
$\beta_{5,4}$&$12.40$&$18.97$&$18.29$&$8.630$&$-187.5$\\
$\beta_5$&$108800$&$2.629\times10^6$&$2.189\times10^7$&$1.046\times10^8$&$1.018\times10^{10}$\\
\hline
\end{tabular}
\end{center}
\caption{\label{qcd6APf} AAPAP predictions for  six-loop QCD $\beta$-function coefficients (inputting $\beta_{5,5}$, w/o~QC)}
\end{table}

\begin{table}
\begin{center}
\begin{tabular}{|c|c|c|c|c|c|}
\hline
$N_C$&2&3&4&5&10\\
\hline
$\beta_{5,0}$&$945100$&$1.076\times10^7$&$6.044\times10^7$&$2.306\times10^8$&$1.475\times10^{10}$\\
$\beta_{5,1}$&$-545000$&$-4.182\times10^6$&$-1.770\times10^7$&$-5.414\times10^7$&$-1.738\times10^9$\\
$\beta_{5,2}$&$106100$&$550200$&$1.754\times10^6$&$4.300\times10^6$&$6.917\times10^7$\\
$\beta_{5,3}$&$-5831$&$-20400$&$-48910$&$-96020$&$-773200$\\
$\beta_{5,4}$&$12.40$&$18.97$&$18.29$&$8.630$&$-187.5$\\
$\beta_5$&$108500$&$2.617\times10^6$&$2.180\times10^7$&$1.042\times10^8$&$1.014\times10^{10}$\\
\hline
\end{tabular}
\end{center}
\caption{\label{qcd6WPf} WAPAP predictions for six-loop QCD $\beta$-function coefficients (inputting $\beta_{5,5}$, w/o~QC)}
\end{table}

In Table~\ref{betful} we show the full ``w/o~QC'' $\beta$-function $\beta_5(N_F)$ for $N_C=3$ and for a range of values of $N_F$, using the different procedures we have described. We see that for lower values of $N_F$, the value of $\beta_5(N_F)$ are fairly uniform whichever procedure is adopted; but for larger $N_F$ the values start to diverge. The maximum spread among AAPAP/WAPAP values for $N_F=3$ is $0.5\%$, widening to $4\%$ when APAP is included. This spread of values is far less than observed at lower loops, which gives us some hope that we are attaining a better accuracy. There was no indication from four or five loops that increasing the range in the APAP or AAPAP case beyond the minimal $[-N_{\rm min},0]$ was a way to maximise the  accuracy, but just for the record: increasing the range in the APAP case gives us roughly a $5\%$ decrease in the prediction of each $\beta(N_F)$ for each step increase in the range, and in the AAPAP case roughly a $1\%$ increase for each step increase; without any clear convergence or reduction in spread of the predictions being apparent. In any case, for our best-guess estimates, our experience in Section~\ref{wapap} would again suggest that we use the WAPAP values (where of course the fitting range is determined for us) with $\beta_{5,5}$ input from Table~\ref{betful}.

\begin{table}
\begin{center}
\begin{tabular}{|c|c|c|c|c|c|c|}
\hline
$N_F$&2&3&4&5&6&8\\
\hline
APAP&$4.375$&$2.614$&$1.748$&$1.760$&$2.663$&$7.348$\\
APAP (inputting $\beta_{5,5}$)&$4.347$&$2.522$&$1.515$&$1.258$&$1.691$&$4.437$\\
AAPAP&$4.442$&$2.619$&$1.525$&$1.032$&$1.014$&$1.881$\\
AAPAP (inputting $\beta_{5,5}$)&$4.444$&$2.629$&$1.548$&$1.083$&$1.112$&$2.175$\\
WAPAP&$4.433$&$2.615$&$1.531$&$1.055$&$1.064$&$2.024$\\
WAPAP (inputting $\beta_{5,5}$)&$4.434$&$2.617$&$1.535$&$1.068$&$1.096$&$2.160$\\
\hline
\end{tabular}
\end{center}
\caption{\label{betful}  Predictions for full six-loop ``w/o~QC'' QCD $\beta$-function $\beta_5(N_F)$ for $N_C=3$ and for given values of $N_F$ ($\times10^6$)}
\end{table}

Next, the ``w.~QC'' results for APAP, AAPAP and WAPAP (with no input) are tabulated in Tables~\ref{qcd6Pa}, \ref{qcd6APa} and \ref{qcd6WPa}; and (with $\beta_{5,5}$ input) in Tables~\ref{qcd6Pfa}, \ref{qcd6APfa} and \ref{qcd6WPfa}.
Here we have included the quartic Casimirs in the four and five loop inputs, and correspondingly we expect these predictions to be closest to the six-loop results with quartic Casimirs included, but probably with potential sextic Casimirs omitted. There is still a fair agreement between the predictions of the first three terms from AAPAP and WAPAP, but it is less pronounced than in the ``w/o~QC'' case.

\begin{table}
\begin{center}
\begin{tabular}{|c|c|c|c|c|c|}
\hline
$N_C$&2&3&4&5&10\\
\hline
$\beta_{5,0}$&$789000$&$9.41\times10^6$&$5.43\times10^7$&$2.11\times10^8$&$1.39\times10^{10}$\\
$\beta_{5,1}$&$-631000$&$-4.53\times10^6$&$-1.87\times10^7$&$-5.67\times10^7$&$-1.79\times10^9$\\
$\beta_{5,2}$&$150000$&$677000$&$2.07\times10^6$&$4.97\times10^6$&$7.79\times10^7$\\
$\beta_{5,3}$&$-7870$&$-25900$&$-60100$&$-115000$&$-879000$\\
$\beta_{5,4}$&$-77.1$&$-297$&$-657$&$-1060$&$-3480$\\
$\beta_{5,5}$&$-10.1$&$-25.5$&$-47.0$&$-69.2$&$-169$\\
$\beta_5$&$27500$&$1.18\times10^6$&$1.50\times10^7$&$8.21\times10^7$&$9.17\times10^{9}$\\
\hline
\end{tabular}
\end{center}
\caption{\label{qcd6Pa} APAP predictions for six-loop QCD $\beta$-function coefficients (no inputs, w.~QC)}
\end{table}

\begin{table}
\begin{center}
\begin{tabular}{|c|c|c|c|c|c|}
\hline
$N_C$&2&3&4&5&10\\
\hline
$\beta_{5,0}$&$857000$&$9.83\times10^6$&$5.58\times10^7$&$2.15\times10^8$&$1.40\times10^{10}$\\
$\beta_{5,1}$&$-636000$&$-4.48\times10^6$&$-1.84\times10^7$&$-5.57\times10^7$&$-1.77\times10^9$\\
$\beta_{5,2}$&$136000$&$624000$&$1.93\times10^6$&$4.68\times10^6$&$7.48\times10^7$\\
$\beta_{5,3}$&$-5790$&$-21400$&$-52800$&$-105000$&$-860000$\\
$\beta_{5,4}$&$15.5$&$-81.0$&$-256$&$-477$&$-2140$\\
$\beta_{5,5}$&$0.149$&$-4.79$&$-11.7$&$-18.6$&$-48.0$\\
$\beta_5$&$15900$&$1.42\times10^6$&$1.65\times10^7$&$8.68\times10^7$&$9.31\times10^{9}$\\
\hline
\end{tabular}
\end{center}
\caption{\label{qcd6APa} AAPAP predictions for six-loop QCD $\beta$-function coefficients (no inputs, w.~QC)}
\end{table}

\begin{table}
\begin{center}
\begin{tabular}{|c|c|c|c|c|c|}
\hline
$N_C$&2&3&4&5&10\\
\hline
$\beta_{5,0}$&$865000$&$9.96\times10^6$&$5.67\times10^7$&$2.18\times10^8$&$1.42\times10^{10}$\\
$\beta_{5,1}$&$-636000$&$-4.48\times10^6$&$-1.85\times10^7$&$-5.60\times10^7$&$-1.78\times10^9$\\
$\beta_{5,2}$&$136000$&$629000$&$1.96\times10^6$&$4.75\times10^6$&$7.58\times10^7$\\
$\beta_{5,3}$&$-5900$&$-21600$&$-52700$&$-104000$&$-849000$\\
$\beta_{5,4}$&$15.5$&$-81.0$&$-256$&$-477$&$-2140$\\
$\beta_{5,5}$&$0.149$&$-4.79$&$-11.7$&$-18.6$&$-48.0$\\
$\beta_5$&$21100$&$1.61\times10^6$&$1.74\times10^7$&$9.00\times10^7$&$9.48\times10^{9}$\\
\hline
\end{tabular}
\end{center}
\caption{\label{qcd6WPa} WAPAP predictions for  six-loop QCD $\beta$-function coefficients (no inputs, w.~QC)}
\end{table}

\begin{table}
\begin{center}
\begin{tabular}{|c|c|c|c|c|c|}
\hline
$N_C$&2&3&4&5&10\\
\hline
$\beta_{5,0}$&$789000$&$9.41\times10^6$&$5.43\times10^7$&$2.11\times10^8$&$1.39\times10^{10}$\\
$\beta_{5,1}$&$-630000$&$-4.53\times10^6$&$-1.87\times10^7$&$-5.67\times10^7$&$-1.79\times10^9$\\
$\beta_{5,2}$&$151000$&$680000$&$2.07\times10^6$&$4.98\times10^6$&$7.79\times10^7$\\
$\beta_{5,3}$&$-7310$&$-24500$&$-57500$&$-111000$&$-870000$\\
$\beta_{5,4}$&$51.0$&$25.1$&$-64.7$&$-189$&$-1350$\\
$\beta_5$&$66000$&$1.28\times10^6$&$1.52\times10^7$&$8.23\times10^7$&$9.17\times10^{9}$\\
\hline
\end{tabular}
\end{center}
\caption{\label{qcd6Pfa} APAP predictions for six-loop QCD $\beta$-function coefficients  (inputting $\beta_{5,5}$, w.~QC)}
\end{table}

\begin{table}
\begin{center}
\begin{tabular}{|c|c|c|c|c|c|}
\hline
$N_C$&2&3&4&5&10\\
\hline
$\beta_{5,0}$&$857000$&$9.83\times10^6$&$5.58\times10^7$&$2.15\times10^8$&$1.40\times10^{10}$\\
$\beta_{5,1}$&$-636000$&$-4.48\times10^6$&$-1.84\times10^7$&$-5.57\times10^7$&$-1.77\times10^9$\\
$\beta_{5,2}$&$136000$&$624000$&$1.93\times10^6$&$4.68\times10^6$&$7.48\times10^7$\\
$\beta_{5,3}$&$-5780$&$-21100$&$-52200$&$-104000$&$-858000$\\
$\beta_{5,4}$&$16.0$&$-17.7$&$-108$&$-239$&$-1530$\\
$\beta_5$&$16100$&$1.44\times10^6$&$1.65\times10^7$&$8.69\times10^7$&$9.31\times10^{9}$\\
\hline
\end{tabular}
\end{center}
\caption{\label{qcd6APfa} AAPAP predictions for six-loop QCD $\beta$-function coefficients (inputting $\beta_{5,5}$, w.~QC)}
\end{table}

\begin{table}
\begin{center}
\begin{tabular}{|c|c|c|c|c|c|}
\hline
$N_C$&2&3&4&5&10\\
\hline
$\beta_{5,0}$&$865000$&$9.96\times10^6$&$5.67\times10^7$&$2.18\times10^8$&$1.42\times10^{10}$\\
$\beta_{5,1}$&$-636000$&$-4.48\times10^6$&$-1.85\times10^7$&$-5.60\times10^7$&$-1.78\times10^9$\\
$\beta_{5,2}$&$136000$&$630000$&$1.96\times10^6$&$4.76\times10^6$&$7.60\times10^7$\\
$\beta_{5,3}$&$-6000$&$-21600$&$-52300$&$-103000$&$-839000$\\
$\beta_{5,4}$&$16.0$&$-17.7$&$-108$&$-239$&$-1530$\\
$\beta_5$&$18500$&$1.62\times10^6$&$1.75\times10^7$&$9.01\times10^7$&$9.49\times10^{9}$\\
\hline
\end{tabular}
\end{center}
\caption{\label{qcd6WPfa} WAPAP predictions for six-loop QCD $\beta$-function coefficients (inputting $\beta_{5,5}$, w.~QC)}
\end{table}

As we did in the ``w/o~QC'' case, we shall now give our best-guess predictions for some of the coefficients. Once again there is wide variation amongst the predictions for $\beta_{5,4}$ and $\beta_{5,5}$ (except where we have constrained them to be the same, as we did for WAPAP and AAPAP). But there is still a fair degree of uniformity among the first four coefficients. We again take the predictions for WAPAP results with $\beta_{5,5}$ input from Table~\ref{qcd6WPfa}. Our most confident predictions are therefore (for $N_C=3$)
\begin{align}
\beta^{\rm{w.~QC}}_{5,0}=&9.96\times 10^6,\quad \beta^{\rm{w.~QC}}_{5,1}=-4.48\times 10^6,\nn
\beta^{\rm{w.~QC}}_{5,2}=&630000,\quad \beta^{\rm{w.~QC}}_{5,3}=-21600.
\label{6resa}
\end{align}

In Table~\ref{betfula} we show the full ``w.~QC'' $\beta$-function $\beta_5(N_F)$ for $N_C=3$ and for a range of values of $N_F$, using the different procedures we have described. We see that for lower values of $N_F$, the value of $\beta$ are somewhat uniform whichever procedure is adopted (though less than the ``w/o~QC'' case); but for larger $N_F$ the values again start to diverge. The maximum spread among AAPAP/WAPAP values for $N_F=3$ is $14\%$, widening to $20\%$ when APAP is included. Once again, this spread of values is reassuringly much less than observed at lower
 loops. As in the  ``w/o~QC'' case, for our best-guess estimates, our previous experience would suggest that we use the WAPAP values with $\beta_{5,5}$ input from Table~\ref{betfula}.

\begin{table}
\begin{center}
\begin{tabular}{|c|c|c|c|c|c|c|}
\hline
$N_F$&2&3&4&5&6&8\\
\hline
APAP&$2.842$&$1.180$&$0.3592$&$0.1814$&$0.4274$&$1.201$\\
APAP (inputting $\beta_{5,5}$)&$2.872$&$1.277$&$0.6035$&$0.7083$&$1.447$&$4.257$\\
AAPAP&$3.194$&$1.422$&$0.5006$&$0.2899$&$0.6471$&$2.468$\\
AAPAP (inputting $\beta_{5,5}$)&$3.200$&$1.441$&$0.5486$&$0.3935$&$0.8475$&$3.068$\\
WAPAP&$3.355$&$1.610$&$0.7225$&$0.5540$&$0.9599$&$2.893$\\
WAPAP (inputting $\beta_{5,5}$)&$3.360$&$1.624$&$0.7586$&$0.6331$&$1.117$&$3.386$\\
\hline
\end{tabular}
\end{center}
\caption{\label{betfula}  Predictions for full six-loop ``w.~QC'' QCD $\beta$-function $\beta_5 (N_F)$ for $N_C=3$ for given values of $N_F$  ($\times10^6$)}
\end{table}

 It is instructive to compare the ``w/o~QC'' predictions in Tables~\ref{qcd6P}-\ref{qcd6WPf} with the corresponding ``w.~QC'' predictions in Tables~\ref{qcd6Pa}-\ref{qcd6WPfa}. There certainly seems to be evidence of convergence between the two sets of predictions, especially for $\beta_{5,0}$ and $\beta_{5,1}$ whose ``w/o~QC'' and ``w.~QC'' predictions differ only by about $10\%$. On the other hand, the ``w/o~QC'' and ``w.~QC'' predictions for $\beta_{5,4}$ and $\beta_{5,5}$ differ considerably, and presumably this accounts for the significant differences in the corresponding results for the full $\beta$-function $\beta_5(N_F)$ in Tables~\ref{betful} and \ref{betfula}. However, if this reflects real similarities and differences in the exact values, it seems to conflict with our experience at lower loops where it was the coefficients of higher powers of $N_F$ which had smaller contributions from quartic Casimirs, and vice-versa. Furthermore, based on our experience at five loops, we would tend to be sceptical of the accuracy of  our ``w.~QC'' predictions, even allowing for the increase in accuracy with loop order which we have noted before.

The usefulness of our ``w/o QC'' predictions is additionally somewhat compromised by our lack of exact knowledge of any ``sextic Casimir'' contributions which are presumably also left out of account by our methods. It is not obvious even how to reliably estimate their effect. It seems likely that they will correspond to primitive graphs which consequently have no relation to lower-loop contributions. Our only natural recourse is to assume that the sextic Casimir contributions will have roughly the same relative magnitude as the quartic Casimir ones, namely around $10\%$ of the full value of the coefficient of each power of $N_F$ in the case of the QCD $\beta$-function, with only moderate dependence on $N_C$ and with indications of a gradual decrease with loop order.

In this context it is of interest to speculate on the composition of the sextic Casimir contribution in terms of $\zeta$-values. Our earlier experience with quartic Casimirs leads us to expect $\zeta$s of the same weight as the remainder of the $\beta$-function; and since the five-loop QCD $\beta$-function contains weight-5 $\zeta$s (e.g. $\zeta_5$), we expect weight-7 $\zeta$s (e.g. $\zeta_7$) at six loops (the weights increasing by two with each loop order). In particular, we do not expect the weight-8 multiple zeta value (MZV) $\zeta_{5,3}$ to appear until seven loops in the sextic Casimir contribution or indeed anywhere in the QCD $\beta$-function.

We pause here to remark an odd feature of the six-loop calculation. We found at five loops that the use of a positive fitting range gave uniformly dire predictions for the QCD $\beta$-function, i.e. results very different from those obtained using a negative range, which were quite accurate. However, this seems no longer the case at six loops. We find the results for the first three coefficients depend very little on whether a positive or negative range is used, especially in the case of APAP; for which the differences between the two ranges are comparable to those between APAP and AAPAP. We shall find similar features in the case of the six-loop quark mass anomalous dimension and in the case of the scalar $\beta$-function. We have taken the decision not to incorporate results from a positive range into our predictions, but it perhaps remains something for future analysis. 

\section{Prediction for six-loop quark mass anomalous dimension in QCD}\label{qcdanom6}
In this Section we derive our predictions for the six-loop quark mass anomalous dimension in QCD. Our experience with this anomalous dimension at five loops was not very encouraging, but on the other hand we have seen some evidence that the accuracy of the predictions increases with loop order, at least in percentage terms, and we shall see even more evidence in Section~\ref{6phi} in the context of scalar $\lambda\phi^4$ theory; therefore it seemed worth persevering. As we did for the six-loop QCD $\beta$-function, we shall present results for both the ``w/o~QC'' and ``w.~QC'' cases. We shall also include the WAPAP predictions, since we found at least some circumstances where this procedure offered an improvement over AAPAP at five loops.
In addition, we shall include the APAP predictions in the ``w/o~QC'' case. At five loops, this procedure gave the best predictions for $\gamma_{4,0}$ in the ``w/o~QC'' case, and comparable predictions to the other procedures in the ``w.~QC'' case, and the six-loop ``w/o~QC'' APAP results are not dissimilar to those for AAPAP; however, the six-loop APAP predictions are very different to AAPAP in the ``w.~QC'' case, so we discard them. We parametrise the six-loop quark mass anomalous dimension in QCD as
 \be
\gamma_5=\gamma_{5,0}+\gamma_{5,1}N_F+\gamma_{5,2}N_F^2+\gamma_{5,3}N_F^3+\gamma_{5,4}N_F^4+\gamma_{5,5}N_F^5.
\label{polygam5}
\ee
The values of $\gamma_{5,5}$ and $\gamma_{5,4}$ have already been derived from large-$N_F$ computations
\cite{pal,derk1,derk2}:
\begin{align}
\gamma_{5,5}=&\tfrac{1}{93312}[-451-560\zeta_3-720\zeta_4+1728\zeta_5]T_F^5C_F,\nn
\gamma_{5,4}=&\tfrac{1}{373248}\Bigl[13824(4C_A-3C_F)(\zeta_3)^2+(112320C_A-25920C_F)\zeta_6\nn
&+1152(-98C_A+45C_F)\zeta_5-(68544C_A+\tfrac{143856}{5}C_F)\zeta_4\nn
&+\tfrac{224}{5}(1444C_A+1191C_F)\zeta_3-\tfrac{1}{10}(198551C_A+154456C_F)\Bigr]T_F^4C_F.
\end{align}
In Tables~\ref{anom6}, \ref{anom6ctd} we show our predictions for $N_C=3$ for the ``w/o~QC'' scenario, and for the cases where we input none, one or two of $\gamma_{5,5}, \gamma_{5,4}$, and using APAP, AAPAP and WAPAP.  We note that, as we did earlier for the five-loop anomalous dimension and the six-loop QCD $\beta$-function, we have discarded WAPAP predictions corresponding to large critical values (bigger than $70$) which tend to be considerably out of line with the others. As before, where necessary in computing $\gamma_5(N_F)$ we have used the corresponding AAPAP values instead. We note here a clear distinction between the predictions of APAP compared with those of AAPAP and WAPAP, but a broad similarity between the results within each procedure for different numbers of inputs, especially for coefficients of low powers of $N_F$. The corresponding WAPAP and AAPAP results are also quite similar, again for coefficients of low powers of $N_F$. Assuming as we did in Section~\ref{6qcd} that the four-loop behaviour should serve as a guide for six loops, and recalling that in Section~\ref{qcdanom} we found that AAPAP with no input was preferable (though WAPAP was unavailable here) we decide to take the average of the AAPAP/WAPAP no-input results in Tables~\ref{anom6}, \ref{anom6ctd}. However, we would tend to place scant credence in any of our results for $\gamma_{5,3}$, $\gamma_{5,4}$ or $\gamma_{5,5}$, in the light of the spread of our predictions and the low accuracy in $\gamma_{4,3}$ seen earlier.   In summary, our best-guess predictions in the ``w/o~QC'' case are 
\be
\gamma_{5,0}^{\rm w/o~QC}=2750,\quad \gamma_{5,1}^{\rm w/o~QC}=-907,\quad \gamma_{5,2}^{\rm w/o~QC}=85.6.
\label{6resAD}
\ee
As in the case of the six-loop QCD $\beta$-function, this selection process is perhaps somewhat questionable and in any case the difference compared with a simple average may well exceed the accuracy we can expect from the Pad\'e procedure.
\begin{table}
\begin{center}
\begin{tabular}{|c|c|c|c|c|c|c|c|}
\hline
&APAP&APAP&APAP&AAPAP&AAPAP&AAPAP\\
&no input&input $\gamma_{5,5}$&input $\gamma_{5,5}, \gamma_{5,4}$&no input&input $\gamma_{5,5}$&input $\gamma_{5,5}, \gamma_{5,4}$\\
\hline
$\gamma_{5,0}$&$2805$&$2805$&$2806$&$2774$&$2774$&$2774$\\
$\gamma_{5,1}$&$-871.1$&$-870.6$&$-867.4$&$-907.3$&$-907.2$&$-906.3$\\
$\gamma_{5,2}$&$94.67$&$95.44$&$98.91$&$85.78$&$85.93$&$86.84$\\
$\gamma_{5,3}$&$-0.6831$&$-0.2440$&$0.8842$&$-0.9525$&$-0.8651$&$-0.5678$\\
$\gamma_{5,4}$&$-0.2499$&$-0.1491$&I/P&$-0.08606$&$-0.06601$&I/P\\
$\gamma_{5,5}$&$-0.008115$&I/P&I/P&$-0.001654$&I/P&I/P\\
\hline
\end{tabular}
\end{center}
\caption{\label{anom6}Predictions  for  six-loop quark mass anomalous dimension coefficients as defined in Eq.~\eqref{polygam5}  for $N_C=3$  (w/o~QC)}
\end{table}

\begin{table}
\begin{center}
\begin{tabular}{|c|c|c|c|}
\hline
&WAPAP&WAPAP&WAPAP\\
&no input&input $\gamma_{5,5}$&input $\gamma_{5,5}, \gamma_{5,4}$\\
\hline
$\gamma_{5,0}$&$2730$&$2744$&$2724$\\
$\gamma_{5,1}$&$-906.2$&$-906.0$&$-905$\\
$\gamma_{5,2}$&$85.39$&$86.17$&$87.82$\\
$\gamma_{5,3}$&$-0.429$&$$&$-0.5486$\\
$\gamma_{5,4}$&$-0.0556$&$-0.02455$&I/P\\
$\gamma_{5,5}$&&I/P&I/P\\
\hline
\end{tabular}
\end{center}
\caption{\label{anom6ctd}Predictions for  six-loop quark mass anomalous dimension coefficients for $N_C=3$; blank entries indicate a failure of the WAPAP procedure (w/o~QC) (continued)}
\end{table}

In Table~\ref{gamful} we show the full six-loop ``w/o~QC'' quark mass anomalous dimension $\gamma_5(N_F)$ for $N_C=3$ and for a range of values of $N_F$, using the different procedures we have described. We see that for $N_F\le5$, the AAPAP and WAPAP estimates of $\gamma_5$ are fairly uniform, but start to diverge rather wildly thereafter. For APAP, we see similar behaviour but with the divergence starting at $N_F=5$. In Section~\ref{qcdanom} we saw that at four loops, AAPAP with no inputs gave good results for $\gamma_3(N_F)$ which improved further as we increased the fitting range. As usual, we might expect this to serve as a guide to six loops. In fact it turns out that at six loops the AAPAP results for $\gamma_5(N_F)$ are quite uniform as we increase the range and vary less wildly than the APAP results as we change the number of inputs. We therefore simply take the minimal-range no-input AAPAP results in Table~\ref{gamful} as our best-guess predictions.

\begin{table}
\begin{center}
\begin{tabular}{|c|c|c|c|c|c|c|}
\hline
$N_F$&1&2&3&4&5&6\\
\hline
APAP&$2026$&$1407$&$874.4$&$312.0$&$-445.7$&$-1612$\\
APAP (inputting $\gamma_{5,5}$)&$2030$&$1441$&$1034$&$796.0$&$714.3$&$771.0$\\
APAP (inputting $\gamma_{5,5}$, $\gamma_{5,4}$)&$2037$&$1473$&$1114$&$965.6$&$1029$&$1305$\\
AAPAP&$1951$&$1293$&$790.9$&$432.5$&$203.9$&$88.01$\\
AAPAP (inputting $\gamma_{5,5}$)&$1952$&$1266$&$797.0$&$447.7$&$236.6$&$151.4$\\
AAPAP (inputting $\gamma_{5,5}$, $\gamma_{5,4}$)&$1954$&$1304$&$818.2$&$492.4$&$319.5$&$292.2$\\
WAPAP&$1909$&$1255$&$763.7$&$428.2$&$240.3$&$189.3$\\
WAPAP (inputting $\gamma_{5,5}$)&$1923$&$1269$&$776.2$&$437.0$&$244.6$&$191.0$\\
WAPAP (inputting $\gamma_{5,5}$, $\gamma_{5,4}$)&$1906$&$1261$&$782.5$&$466.0$&$304.8$&$291.7$\\
\hline
\end{tabular}
\end{center}
\caption{\label{gamful}  Predictions for full six-loop ``w/o~QC'' quark mass anomalous dimension $\gamma_5(N_F)$ for $N_C=3$ and for given values of $N_F$  }
\end{table}

In Table~\ref{anom6a} we show the six-loop predictions for AAPAP and WAPAP in the ``w.~QC'' case. Once again, we see a broad agreement between the results within AAPAP and WAPAP separately for different numbers of inputs, especially for coefficients of low powers of $N_F$. Our best-guess predictions in the ``w.~QC'' case are (taking the averages of  the no-input AAPAP and WAPAP values for $\gamma_{5,0}$, $\gamma_{5,1}$ and $\gamma_{5,2}$ in Table~\ref{anom6a})
\be
\gamma_{5,0}^{\rm w.~QC}=3530,\quad \gamma_{5,1}^{\rm w.~QC}=-1150,\quad \gamma_{5,2}^{\rm w.~QC}=99.4.
\label{6resADa}
\ee

\begin{table}
\begin{center}
\begin{tabular}{|c|c|c|c|c|c|c|}
\hline
&AAPAP&AAPAP&AAPAP&WAPAP&WAPAP&WAPAP\\
&no input&input $\gamma_{5,5}$&input $\gamma_{5,5}, \gamma_{5,4}$&no input&input $\gamma_{5,5}$&input $\gamma_{5,5}, \gamma_{5,4}$\\
\hline
$\gamma_{5,0}$&$3591$&$3591$&$3591$&$3476$&$3467$&$3433$\\
$\gamma_{5,1}$&$-1150$&$-1150$&$-1149$&$-1144$&$-1144$&$-1143$\\
$\gamma_{5,2}$&$100.5$&$100.7$&$101.7$&$98.28$&$99.17$&$100.9$\\
$\gamma_{5,3}$&$-0.8352$&$-0.7441$&$-0.4329$&&&$-0.5232$\\
$\gamma_{5,4}$&$-0.08831$&$-0.06740$&I/P&$-0.05175$&$-0.02284$&I/P\\
$\gamma_{5,5}$&$-0.001723$&I/P&I/P&&I/P&I/P\\
\hline
\end{tabular}
\end{center}
\caption{\label{anom6a}Predictions for six-loop quark mass anomalous dimension coefficients  for $N_C=3$; blank entries indicate a failure of the WAPAP procedure (w.~QC)}
\end{table}

In Table~\ref{gamfula} we show the full six-loop ``w.~QC'' quark mass anomalous dimension $\gamma_5(N_F)$ for $N_C=3$ and for a range of values of $N_F$, using the different procedures we have described. We see that for $N_F\le4$, the AAPAP and WAPAP estimates of $\gamma_5$ are fairly uniform (though with a fairly wide separation between AAPAP and WAPAP), but start to diverge significantly thereafter. Once again we suggest taking the minimal-range no-input AAPAP results in Table~\ref{gamfula} as our best-guess predictions.

\begin{table}
\begin{center}
\begin{tabular}{|c|c|c|c|c|c|c|}
\hline
$N_F$&1&2&3&4&5&6\\
\hline
AAPAP&$2541$&$1686$&$1022$&$543.3$&$243.1$&$114.6$\\
AAPAP (inputting $\gamma_{5,5}$)&$2541$&$1686$&$1021$&$536.6$&$222.1$&$66.39$\\
AAPAP (inputting $\gamma_{5,5}$, $\gamma_{5,4}$)&$2543$&$1695$&$1044$&$583.3$&$308.9$&$213.8$\\
WAPAP&$2429$&$1573$&$900.5$&$404.0$&$74.07$&$-100.1$\\
WAPAP (inputting $\gamma_{5,5}$)&$2422$&$1569$&$905.3$&$423.7$&$118.1$&$-18.52$\\
WAPAP (inputting $\gamma_{5,5}$, $\gamma_{5,4}$)&$2391$&$1547$&$896.3$&$434.1$&$154.1$&$49.10$\\
\hline
\end{tabular}
\end{center}
\caption{\label{gamfula} Predictions for full six-loop ``w.~QC'' quark mass anomalous dimension $\gamma_5(N_F)$ for $N_C=3$ and for given values of $N_F$  }
\end{table}

As in the case of the QCD $\beta$-function, it is interesting to compare the ``w/o~QC'' predictions in Table~\ref{anom6} with the corresponding ``w.~QC'' predictions in Table~\ref{anom6a}. In fact, the behaviour seems somewhat the reverse of the previous case. There is no evidence of convergence between the two sets of predictions, even for $\gamma_{5,0}$ and $\gamma_{5,1}$. On the other hand, the differences between the corresponding results for the full mass anomalous dimension $\gamma_5(N_F)$ in Tables~\ref{gamful} and \ref{gamfula} are considerably less than we saw for the $\beta$-function. Once again, this seems to conflict with our experience at five loops where there was little difference between the ``w/o~QC'' and ``w.~QC'' predictions of the low-$N_F$ coefficients $\gamma_{4,0}$ and $\gamma_{4,1}$, in line with the small contributions from quartic Casimirs to the anomalous dimension. The odd behaviour might simply be a consequence of the ``w. QC'' predictions being rather poor at six loops; but the five-loop ``w. QC'' anomalous dimension predictions were quite good, and moreover we have come to expect an increase in accuracy with loop order. We do not have enough information to resolve this puzzle.

Once again, we expect that our predictions will leave out of account any potential sextic Casimir contributions, but again we can at least speculate on their composition in terms of $\zeta$-values. We expect $\zeta$s of the same weight as the remainder of the quark mass anomalous dimension; and since the five-loop quark mass anomalous dimension contains weight-7 $\zeta$s, we expect weight-9 $\zeta$s at six loops. Therefore we might anticipate the weight-8 multiple zeta value (MZV) $\zeta_{5,3}$ to appear in both the the sextic Casimir contribution and the remainder of the anomalous dimension at six loops (which is reminiscent of the $\zeta$ structure of the $\lambda\phi^4$ $\beta$-function\cite{panz}).

Finally, we mention that as in the case of the six-loop QCD $\beta$-function, we found some odd behaviour when we investigated the use of a positive fitting range. We focussed on the ``w/o~QC'' case, since there was some evidence at five loops that a positive range worked quite well here. We found that the positive-range AAPAP predictions were of the same order as the negative-range APAP and AAPAP predictions (in fact the differences were comparable to, but mostly somewhat greater than, those between APAP and AAPAP); whereas the APAP results for positive range were wildly different, except for $\gamma_{5,0}$. Nevertheless, as in the case of the six-loop QCD $\beta$-function, we have taken the decision to discount these positive-range results. 

\section{Scalar $O(N)$ $\lambda\phi^4$ theory}\label{6phi}

In this Section we explore the use of Asymptotic Pad\'e Approximants in $O(N)$ scalar $\lambda\phi^4$ theory. We have tested the use of WAPAP but in the case of $\lambda\phi^4$ theory we have found no evidence that the use of the weight criterion provides more accurate results than the straightforward AAPAP predictions; so in this Section and the next, we focus on APAP and AAPAP. In fact, the APAP and AAPAP methods were developed  in Ref.~\cite{sam1} and tested there by comparing four and five-loop predictions in scalar $\lambda\phi^4$ theory with known results\cite{kaztar,chet4,chet5,chet6}. A prediction for the six-loop $\beta$-function was also given in Ref.~\cite{sam1}, and later compared with the newly-computed six-loop $\lambda\phi^4$ $\beta$-function in Ref.~\cite{panz}. In this Section we shall revisit these calculations and also continue with a ``postdiction'' for the seven-loop $\beta$-function which was computed in Ref.~\cite{schnetz}, in order to get some insight into how accurate our later eight-loop predictions might be. The Lagrangian is given by 
\be
{\cal L}=\tfrac12\pa_{\mu}\phi^i\pa^{\mu}\phi^i-\tfrac{16\pi^2}{4!}\lambda(\phi^i\phi^i)^2,\quad 1\le i \le N
\ee
(ignoring, for simplicity, the possibility of a mass term). We have included the factor of $16\pi^2$ to absorb $\pi$-dependent factors which in the earlier QCD calculations were accounted for by our definitions of the couplings $a_s$ and $a$. The quantities $N$ (multiplicity of scalar fields) and $N_F$ (number of quark flavours) play very similar roles in our Pad\'e approximation procedures and, as we remarked in Section~\ref{6qcd}, we might expect qualitatively similar behaviour; nevertheless to avoid confusion we have retained the usual notations.
The $\beta$-function is written
\be
\beta^{\lambda}=\sum_{n=0} \beta_n^{\lambda}\lambda^{n+2}
\ee
and the one, two and three-loop $\beta$-functions are given by
\begin{align} 
\beta^{\lambda}_0=&\tfrac16(N+8),\quad \beta^{\lambda}_1=-\tfrac16(3N+14),\nn
\beta^{\lambda}_2=&\tfrac{1}{432}[96(5N+22)\zeta_3+33N^2+922N+2960].
\end{align}
We parametrise the four, five, six and seven-loop $\beta$-functions as
\begin{align}
\beta^{\lambda}_3=&\beta^{\lambda}_{3,0}+\beta^{\lambda}_{3,1}N+\beta^{\lambda}_{3,2}N^2+\beta^{\lambda}_{3,3}N^3,\nn
\beta^{\lambda}_4=&\beta^{\lambda}_{4,0}+\beta^{\lambda}_{4,1}N+\beta^{\lambda}_{4,2}N^2+\beta^{\lambda}_{4,3}N^3+\beta^{\lambda}_{4,4}N^4,\nn
\beta^{\lambda}_5=&\beta^{\lambda}_{5,0}+\beta^{\lambda}_{5,1}N+\beta^{\lambda}_{5,2}N^2+\beta^{\lambda}_{5,3}N^3+\beta^{\lambda}_{5,4}N^4+\beta^{\lambda}_{5,5}N^5,\nn
\beta^{\lambda}_6=&\beta^{\lambda}_{6,0}+\beta^{\lambda}_{6,1}N+\beta^{\lambda}_{6,2}N^2+\beta^{\lambda}_{6,3}N^3+\beta^{\lambda}_{6,4}N^4+\beta^{\lambda}_{6,5}N^5+\beta^{\lambda}_{6,6}N^6.
\label{lamb4}
\end{align}
 The results were derived in Refs.~\cite{brez}, \cite{kaztar,ditt}, \cite{chet4,chet5,chet6}, \cite{panz} and \cite{schnetz} respectively, but we do not present them in detail here, though we give numerical values later in Tables~\ref{phi4}, \ref{phi5}, \ref{phi6} and \ref{phi7}.
In extending to seven loops, in the spirit of ``postdiction'' we pretend we do not know the seven-loop $\beta$-function (with the exception of the leading two powers of $N$ which are known from large-$N$ calculations\cite{bgk,jag1}); but now we are ``allowed'' to use the known six-loop $\beta$-function $\beta^{\lambda}_5$. Therefore we can now compute
\be
\delta_5=\frac{\beta_5^{\lambda\rm{PAP}}-\beta^{\lambda}_5}{\beta^{\lambda}_5}
\ee 
in addition to $\delta_2$, $\delta_3$, and $\delta_4$ in Eq.~\eqref{deltas}, and so $\Acal$ and $a+b$ are even more overdetermined than in the six-loop case. Once again, our proposal is that since we are exploiting an asymptotic property in Eq.~\eqref{delasy}, we should use the highest-order information we have, in this case $\delta_4$ and $\delta_5$; and at six loops, we continue to use the pair $\{\delta_3,\delta_4\}$ to determine $\Acal$ and $a+b$, as we did in the QCD case in Section~\ref{6qcd}. But now we should be a little more careful than in Section~\ref{6qcd}. As we explained in that Section, the accuracies for the coefficients of the first two or three powers of $N$ in the scalar $\beta$-function do seem to be best for the choice $\{\delta_3,\delta_4\}$ at six loops and $\{\delta_4,\delta_5\}$ at seven loops, for both AAPAP and APAP; but this is certainly not true for the coefficients of higher powers of $N$. This makes it difficult to make general statements about relative accuracies. However in the case of QCD, we are interested in low values of $N_F$, which corresponds to low values of $N$ in the scalar case. If we focus on the values of the full $\beta$-function $\beta^{\lambda}(N)$ for the value $N\le6$, for instance,  there is evidence (as we saw in Section~\ref{6qcd}) that this strategy for picking pairs of $\delta$s gives more accurate results for $O(N)$ theory, at both six and seven loops (of course this still begs the question of whether it is legitimate to extrapolate from $N\le6$ in $O(N)$ to $N_F\le6$ in QCD). Now we would like to examine what we can say about the accuracy of $\beta^{\lambda}(N)$ for higher values of $N$. The accuracies found for $\beta^{\lambda}(N)$ using AAPAP are always positive for any $N$ and the choice of $\delta_3,\delta_4$ at six loops and $\delta_4,\delta_5$ at seven loops turns out to be almost always significantly better than any other choice. In the case of APAP, the accuracies of $\beta^{\lambda}(N)$ for increasing $N$ usually end up (at least at seven loops) by decreasing and changing sign at some value of $N$, and at different values of $N$ for different choices of pairs of $\delta$; it is therefore difficult, if not impossible, to make a general statement about which of these choices gives the best overall accuracy. Certainly there is no choice of a different pair of $\delta$ (than $\{\delta_4,\delta_5\}$) which delivers a better accuracy for every value of $N$ (as we have already seen, our current choices are superior to the rest for low values of $N$) but $\delta_2,\delta_5$ and $\delta_3,\delta_5$ do give an improvement for particular, larger values of $N$. Nevertheless, aside from this ``cherry-picking'' of particular values of $N$, our original strategy does yield the most accurate results overall, for both APAP and AAPAP. One might also entertain the heretical idea that at these higher loop orders one would do better to abandon the idea of asymptotic corrections; but in fact any of the choices of pairs of $\delta$ is dramatically better than the basic uncorrected PAP. Of course our real interest is at eight loops where we cannot yet compare relative accuracies; but what we can say (to anticipate our results in the next Section) is that the differences in the eight-loop predictions obtained for the full $\beta$-function $\beta_7^{\lambda}(N)$ using different pairs of $\delta$ are very small and no bigger than the differences resulting from our other choices such as APAP versus AAPAP (though again, there is a significant difference between using any of these pairs of $\delta$, and returning to the PAP case without asymptotic corrections).

In Tables~\ref{phi4}, \ref{phi5}, \ref{phi6} and \ref{phi7} we display the numerical values of the exact results, and the accuracies produced by use of APAP and AAPAP for various choices of  input coefficients; for four, five, six and seven loops respectively. As mentioned earlier, the exact results may be extracted from Ref.~\cite{kaztar} at four loops, Refs.~\cite{chet4,chet5,chet6} at five loops, Ref.~\cite{panz} at six loops and Ref.~\cite{schnetz} at seven loops. In each case, at $L$ loops we have used the minimum fitting range $N_{\rm min}=L-1$ which can be applied for zero, one, or two  coefficients, as described in Section~\ref{qcd5}; we have experimented with larger fitting ranges, but found no consistent improvement which might serve as a guide to strategy (the error often decreases monotonically with the size of range, but with no way to predict at what point it passes through zero).  We observe that the signs here do not alternate (though the pattern is broken for high powers of $N$, as indeed we saw for the QCD $\beta$-function). Accordingly, we have used a positive range of values of $N$ for both APAP and AAPAP\footnote{We note here that we have quite closely reproduced the four-loop and five-loop APAP and AAPAP predictions given in Tables I and III of Ref.~\cite{sam1}, though with the occasional difference in the final significant figure quoted (taking due account of their use of the range of $[0,4]$ in both cases, which differs from our choice in the four-loop case). Similarly, we have not managed to reproduce exactly the six-loop $\lambda\phi^4$ predictions given in Ref.~\cite{sam1}; namely, for the full six-loop $\beta$-function they obtain $-11828$, $-17687$, $-24958$, $-33802$, $-44330$ for $N=0,1,2,3,4$, respectively, whereas we get $-11862$, $-17616$, $-24690$, $-33213$, $-43318$ using APAP and $-11673$, $-17449$, $-24627$, $-33344$ and $-43736$ using AAPAP. Their numbers are certainly closer to our APAP for $N=0,1$ but then move towards AAPAP for larger $N$. The accuracy of these six-loop Pad\'e predictions as given in Ref.~\cite{sam1} was assessed in Table V of Ref.~\cite{panz} in the light of their exact results as around $1.5-2\%$; but our own predictions are more accurate, as we shall see shortly in Table~\ref{phi6acca}. We do agree more closely (after allowing for a relative factor) with the six-loop APAP results for the full $\beta$-function given in Table 1 of Ref.~\cite{chish}, and agree exactly with the errors assigned to these predictions in Table V of Ref.~\cite{panz}, as may be seen in Table~\ref{phi6acca}.}. We observe a propensity for APAP to be more accurate than AAPAP for the second and maybe third coefficient at five loops and seven loops (though there is also an anomalously good APAP prediction for $\beta^{\lambda}_{5,3}$ at six loops). We also observe that the overall accuracies for APAP are better at five and seven loops than at six loops, especially for the leading coefficient. One might wonder if there is a tendency for properties to skip a loop order. Despite these slight anomalies, the dramatic increase in accuracy provided by AAPAP for the first coefficient probably still makes it the method of choice for $\lambda\phi^4$ theory compared with APAP. There is a contrast here with the five-loop QCD $\beta$ function where APAP performed considerably better than AAPAP for the first coefficient. As we commented earlier, it was the observation in Ref.~\cite{sam1} that AAPAP gave better results than APAP for $\lambda\phi^4$ theory at four loops which prompted the use of AAPAP for QCD predictions at four loops in Ref.~\cite{sam1}; and the same averaging procedure for $\Acal$ also underlies the definition of WAPAP in Ref.~\cite{wap1}, as we saw in Section~\ref{wapap}.
The extra information we now have from explicit higher loop calculations seems to support this policy.

\begin{table}
\begin{center}
\begin{tabular}{|c|c||c|c|c|c|c|c|}
\hline
&Exact&APAP&AAPAP&APAP&AAPAP&APAP&AAPAP\\
\hline
$\beta^{\lambda}_{3,0}$&$-100.5$&$4.7\%$&$3.9\%$&$4.7\%$&$3.9\%$&$5.1\%$&$4.4\%$\\
$\beta^{\lambda}_{3,1}$&$-33.28$&$2.2\%$&$3.9\%$&$1.8\%$&$3.7\%$&$-1.9\%$&$-0.91\%$\\
$\beta^{\lambda}_{3,2}$&$-2.059$&$-26\%$&$-23\%$&$-20\%$&$-20\%$&I/P&I/P\\
$\beta^{\lambda}_{3,3}$&$\tfrac{5}{7776}$&$-4300\%$&$-1700\%$&I/P&I/P&I/P&I/P\\
\hline
\end{tabular}
\end{center}
\caption{\label{phi4} Exact results and accuracies of APAP and AAPAP predictions for four-loop scalar $\beta$-function coefficients as defined in Eq.~\eqref{lamb4} }
\end{table}

\begin{table}
\begin{center}
\begin{tabular}{|c|c||c|c|c|c|c|c|}
\hline
&Exact&APAP&AAPAP&APAP&AAPAP&APAP&AAPAP\\
\hline
$\beta^{\lambda}_{4,0}$&$1002$&$-1.1\%$&$0.0095\%$&$-1.1\%$&$0.010\%$&$-1.1\%$&$0.016\%$\\
$\beta^{\lambda}_{4,1}$&$385.6$&$0.55\%$&$1.3\%$&$0.43\%$&$1.3\%$&$0.12\%$&$1.2\%$\\
$\beta^{\lambda}_{4,2}$&$36.12$&$16\%$&$6.7\%$&$18\%$&$7.5\%$&$20\%$&$8.3\%$\\
$\beta^{\lambda}_{4,3}$&$0.5764$&$70\%$&$28\%$&$25\%$&$8.7\%$&I/P&I/P\\
$\beta^{\lambda}_{4,4}$&$-0.001287$&$2562\%$&$1082\%$&I/P&I/P&I/P&I/P\\
\hline
\end{tabular}
\end{center}
\caption{\label{phi5}Exact results and accuracies of APAP and AAPAP predictions for five-loop scalar $\beta$-function coefficients as defined in Eq.~\eqref{lamb4}  }
\end{table}

\begin{table}
\begin{center}
\begin{tabular}{|c|c||c|c|c|c|c|c|}
\hline
&Exact&APAP&AAPAP&APAP&AAPAP&APAP&AAPAP\\
\hline
$\beta^{\lambda}_{5,0}$&$-11660$&$1.8\%$&$0.14\%$&$1.8\%$&$0.14\%$&$1.8\%$&$0.14\%$\\
$\beta^{\lambda}_{5,1}$&$-5083$&$1.0\%$&$0.68\%$&$1.0\%$&$0.68\%$&$1.1\%$&$0.71\%$\\
$\beta^{\lambda}_{5_2}$&$-626.3$&$-4.3\%$&$1.4\%$&$-4.4\%$&$1.5\%$&$-5.4\%$&$1.3\%$\\
$\beta^{\lambda}_{5,3}$&$-21.24$&$-8.4\%$&$4.1\%$&$-6.1\%$&$2.2\%$&$2.6$&4.7\\
$\beta^{\lambda}_{5,4}$&$-0.05667$&$520\%$&$-64\%$&$330\%$&$95\%$&I/P&I/P\\
$\beta^{\lambda}_{5,5}$&$0.0001051$&$8400\%$&$-6900\%$&I/P&I/P&I/P&I/P\\
\hline
\end{tabular}
\end{center}
\caption{\label{phi6}Exact results and accuracies of APAP and AAPAP predictions for six-loop scalar $\beta$-function coefficients as defined in Eq.~\eqref{lamb4} }
\end{table}

\begin{table}
\begin{center}
\begin{tabular}{|c|c||c|c|c|c|c|c|}
\hline
&Exact&APAP&AAPAP&APAP&AAPAP&APAP&AAPAP\\
\hline
$\beta^{\lambda}_{6,0}$&$303900$&$0.12\%$&$-0.061\%$&$0.12\%$&$-0.061\%$&$0.12\%$&$-0.061\%$\\
$\beta^{\lambda}_{6,1}$&$147400$&$0.32\%$&$0.46\%$&$0.30\%$&$0.40\%$&$0.32\%$&$0.40\%$\\
$\beta^{\lambda}_{6,2}$&$22230$&$0.72\%$&$0.27\%$&$0.97\%$&$1.1\%$&$0.77\%$&$1.1\%$\\
$\beta^{\lambda}_{6,3}$&$1151$&$6.2\%$&$14\%$&$2.7\%$&$2.0\%$&$4.5\%$&$2.2\%$\\
$\beta^{\lambda}_{6,4}$&$14.355$&$-90\%$&$-290\%$&$-3.3\%$&$4.0\%$&$-31\%$&$0.56\%$\\
$\beta^{\lambda}_{6,5}$&$-0.0194$&$-8065\%$&$-32000\%$&$1400\%$&$170\%$&I/P&I/P\\
$\beta^{\lambda}_{6,6}$&$5.935\times10^{-6}$&$-0.17\times10^7\%$&$-0.58\times10^7\%$&I/P&I/P&I/P&I/P\\
\hline
\end{tabular}
\end{center}
\caption{\label{phi7}Exact results and accuracies of APAP and AAPAP predictions for seven-loop scalar $\beta$-function coefficients as defined in Eq.~\eqref{lamb4}   }
\end{table}

\begin{table}
\begin{center}
\begin{tabular}{|c|c|c|c|c|c|c|c|c|c|c|}
\hline
$N$&2&3&4&5&6\\
\hline
APAP &$2.4\%$&$1.3\%$&$0.30\%$&$-0.51\%$&$-1.1\%$\\
AAPAP&$2.7\%$&$1.8\%$&$0.75\%$&$-0.16\%$&$-0.93\%$\\
\hline
\hline
$N$&8&10&12&16&20\\
\hline
APAP &$-2.0\%$&$-2.3\%$&$-2.2\%$&$-1.1\%$&$0.78\%$\\
AAPAP&$-2.1\%$&$-2.8\%$&$-3.0\%$&$-2.7\%$&$-1.5\%$\\
\hline
\end{tabular}
\end{center}
\caption{\label{phi4acca} Accuracies of  predictions for full four-loop scalar $\beta$-function $\beta^{\lambda}_3(N)$ for given values of $N$}
\end{table}

\begin{table}
\begin{center}
\begin{tabular}{|c|c|c|c|c|c|c|c|c|c|c|}
\hline
$N$&2&3&4&5&6&8&10&12&16&20\\
\hline
APAP &$1.0\%$&$2.3\%$&$3.5\%$&$4.5\%$&$5.5\%$&$6.9\%$&$7.7\%$&$7.9\%$&$7.2\%$&$5.3\%$\\
AAPAP&$1.1\%$&$1.6\%$&$2.1\%$&$2.8\%$&$3.5\%$&$4.6\%$&$5.3\%$&$5.7\%$&$5.7\%$&$4.9\%$\\
\hline
\end{tabular}
\end{center}
\caption{\label{phi5acca} Accuracies of  predictions for full five-loop scalar $\beta$-function $\beta^{\lambda}_4(N)$ for given values of $N$ }
\end{table}

\begin{table}
\begin{center}
\begin{tabular}{|c|c|c|c|c|c|c|c|c|c|c|}
\hline
$N$&2&3&4&5&6\\
\hline
APAP &$0.78\%$&$0.28\%$&$-0.15\%$&$-0.5\%$&$-0.75\%$\\
AAPAP&$0.52\%$&$0.67\%$&$0.81\%$&$0.94\%$&$0.80\%$\\
\hline
\hline
$N$&8&10&12&16&20\\
\hline
APAP &$-1.0\%$&$-1.0\%$&$-0.81\%$&$-0.39\%$&$0.94\%$\\
AAPAP&$0.62\%$&$0.54\%$&$0.58\%$&$0.93\%$&$1.6\%$\\
\hline
\end{tabular}
\end{center}
\caption{\label{phi6acca} Accuracies of predictions for full six-loop scalar $\beta$-function $\beta^{\lambda}_5(N)$ for given values of $N$}
\end{table}

\begin{table}
\begin{center}
\begin{tabular}{|c|c|c|c|c|c|c|c|c|c|c|c|}
\hline
$N$&2&3&4&5&6\\
\hline
APAP &$0.34\%$&$0.45\%$&$0.54\%$&$0.62\%$&$0.66\%$\\
AAPAP&$0.32\%$&$0.46\%$&$0.59\%$&$0.70\%$&$0.80\%$\\
\hline
\hline
$N$&8&10&12&16&20\\
\hline
APAP &$0.65\%$&$0.52\%$&$0.28\%$&$-0.42\%$&$-1.3\%$\\
AAPAP&$0.90\%$&$0.94\%$&$0.93\%$&$0.78\%$&$0.50\%$\\
\hline
\end{tabular}
\end{center}
\caption{\label{phi7acca} Accuracies of predictions for full seven-loop scalar $\beta$-function $\beta^{\lambda}_6(N)$ for given values of $N$}
\end{table}

Now we turn to considering estimates for the full $\beta$-function $\beta^{\lambda}(N)$, as indeed we did in earlier Sections. In Section~\ref{qcd5} we estimated the coefficients of powers of $N_F$ in the polynomial for $\beta_4$ in Eq.~\eqref{poly4} by fitting Pad\'e estimates of $\beta_4(N_F)$ for a negative range of values of $N_F$, arguing that this was more accurate due to the alternating signs of the coefficients. We then obtained our estimates of $\beta_4(N_F)$ for positive values of $N_F$ by evaluating our estimated polynomial at those values. As we commented there, an alternative option would have been simply to compute Pad\'e estimates directly at those positive values; it could have been that this gave better results for certain limited ranges of $N_F$.  However, an explicit computation shows that this procedure leads to less accurate predictions. But now the situation is different. In the $\lambda\phi^4$ case, we are estimating the polynomial for $\beta^{\lambda}(N)$ by fitting over a positive range of $N$. If we want an estimate for $\beta^{\lambda}(N)$ for a positive value of $N$, there is no reason to use anything other than our standard Pad\'e estimate for $\beta^{\lambda}(N)$, since now these estimates are precisely our inputs for the fitting process. The alternative procedure of going through the fitting process and then inserting the required positive values of $N$ into our estimated polynomial for $\beta^{\lambda}(N)$ will be more laborious and lead to results which are either identical, or less accurate. At the risk of labouring the point, let us explain this in more detail, taking  the four-loop case as a concrete example, with $\beta^{\lambda}_3$ parametrised as in Eq.~\eqref{lamb4}. In the no-input case, we have considered a fit to the cubic polynomial over the range $[0,3]$, where we input four Pad\'e estimates for the $\beta$-function, for the values $N=0,1,2,3$. The fit determines the four unknowns $\beta^{\lambda}_{3,0}\ldots\beta^{\lambda}_{3,3}$. It is then clear that inserting the four values $N=0,1,2,3$
into our estimated $\beta$-function polynomial $\beta_3^{\lambda}$ in Eq.~\eqref{lamb4} will exactly reproduce the input Pad\'e estimates. Furthermore, if one fits over a wider range $[0,N_{\rm max}]$ where $N_{\rm max}>3$, and therefore inputs estimates for the range $N=0,\ldots N_{\rm max}$, the values  obtained by inputting $N=0,\ldots N_{\rm max}$ into the resulting Pad\'e estimate polynomial are still very close to the corresponding input estimates; as may be seen by an explicit computation. It may seem surprising that we can get a close approximation to more than four values using a cubic polynomial with four parameters, but presumably this is because there is an underlying exact polynomial determining the exact values corresponding to the estimates. Finally, if one fits over a range $[0,N_{\rm max}]$ where $N_{\rm max}>3$, and then inputs values $N>N_{\rm max}$, these estimates start to diverge more widely from the original Pad\'e estimates for $\beta^{\lambda}_3(N)$, which for $N>N_{\rm max}$ are not being used as inputs. Therefore, as we claimed, the original Pad\'e estimates for $\beta^{\lambda}_3(N)$ are the most accurate available for general $N$. Of course this applies equally either when including known input coefficients, or at higher loops, or whether we apply the APAP or AAPAP procedure. We shall continue to use the same nomenclature of APAP/AAPAP whether we estimate the coefficents in the polynomial or simply the values $\beta^{\lambda}(N)$; just to be clear, in this latter case we obtain the $\Acal$ used to compute the estimate for $\beta^{\lambda}(N)$ in the APAP scenario from the single value $N$, and in the AAPAP scenario by averaging the values of $\Acal$ obtained over the range $[0,N]$.

The results for the errors in $\beta^{\lambda}(N)$ for selected values of $N$ at four, five, six and seven loops are shown in Tables~\ref{phi4acca}, \ref{phi5acca}, \ref{phi6acca} and \ref{phi7acca} respectively. As we see, this is a slightly cumbersome way to present the information, and one might feel that it is more efficient to present the errors in the predicted coefficients as in Tables~\ref{phi4}, \ref{phi5}, \ref{phi6} and \ref{phi7}, and accept the slight loss of accuracy incurred when using the polynomial with estimated coefficients to estimate $\beta^{\lambda}(N)$ for larger values of $N$ (of course a sensible strategy would be to use a fitting range extending to the highest value of $N$ one expects to encounter). We note that AAPAP is an improvement over APAP in only about $50\%$ of cases.

Notwithstanding the arguable redundancy of this presentation, there are several interesting observations arising from Tables~\ref{phi4acca}, \ref{phi5acca}, \ref{phi6acca} and \ref{phi7acca}. Firstly, the errors in $\beta^{\lambda}(N)$ at even loops appear to be decreasing with loop order, as do those at odd loops (for a given $N$ and a given choice of APAP or AAPAP); even though the five-loop and seven-loop errors are typically bigger than the four-loop and six-loop errors, respectively.  It is also interesting to observe that both the APAP and AAPAP errors at five and seven loops rise then fall with increasing $N$, while the APAP and AAPAP errors for four loops, and the APAP errors for six loops, fall then rise. The AAPAP errors at six loops buck the trend by rising, falling, and rising again. Nevertheless, all this confirms the impression of an alternation between odd and even loop behaviour.

Secondly, the errors are all positive for low enough $N$, but almost all the APAP errors and most of the AAPAP errors eventually turn negative with increasing $N$. This is another example of the phenomenon mentioned earlier, where there can be an exceptionally high but somewhat fortuitous accuracy for particular values of $N$ or $N_C$. 

Finally, the accuracies we obtain here for $\beta^{\lambda}(N)$ are typically much better than the results obtained for $\beta(N_F)$ in QCD. 
We didn't list the QCD results obtained by direct estimate, but the errors were of the same order as the APAP no-input errors given in Tables~\ref{fullbet4}, \ref{fullbet} (which were obtained using the polynomial estimates). We don't completely understand the reason for this difference; however if one tries to estimate $\beta^{\lambda}(N)$ for negative values of $N$ one finds the errors are similarly large.

In view of our experience in the QCD $\beta$-function and anomalous dimension cases (where there were hints that at higher loops a positive fitting range might sometimes be competitive in accuracy with the usual negative range) we experimented with a negative range in the scalar case. The results were almost invariably worse for four, five, six and seven loops, and often considerably so; and it was difficult to detect any trend, either with increasing loop order, or between the performances of APAP and AAPAP. Nevertheless we also considered a positive range when we proceeded to eight loops, as we shall describe in due course.

At this point, as promised earlier, we shall return to consider the arguments presented in Ref.~\cite{sam1} for using AAPAP rather than APAP and for a similar averaging procedure to be used as part of the definition of WAPAP in Ref.~\cite{wap1}. The argument appears to rest simply on the four-loop results for $\lambda\phi^4$ theory which we have reproduced in Table~\ref{phi4} (as we mentioned earlier, we have not been able to reproduce the numbers in Table 1 of Ref.~\cite{sam1} exactly, but this does not affect the discussion). Certainly the first coefficient (in the one-input scenario as used in Ref.~\cite{sam1}) is given more accurately within AAPAP, but the other two are given more accurately using APAP. Furthermore we see in Table~\ref{phi4acca} that the full four-loop $\beta$-function $\beta^{\lambda}_3(N)$ is given better by APAP for low values of $N$. Ref.~\cite{sam1} presents a graph in Fig.~1a showing that the full AAPAP $\beta$-function very closely follows the true value as a function of $N$; but it does not show the APAP version which would be even closer. In hindsight there seems to have been a certain amount of luck in chancing upon the particular scenario in Table~\ref{fullbet4} which gave the best accuracy for four-loop QCD; but it is not clear how we would know in advance which would be the best five-loop choice in Table~\ref{fullbet}.

\section{Predictions for eight-loop $\lambda\phi^4$ theory $\beta$-function}\label{8phi}
In this Section we shall present predictions for the eight-loop $\beta$-function in $O(N)$ $\lambda\phi^4$ theory using both the APAP and AAPAP procedures and inputting various known eight-loop coefficients, before selecting what we shall argue are likely to be the most accurate values.
We parametrise the eight-loop $\beta$-function as
\be
\beta^{\lambda}_7=\beta^{\lambda}_{7,0}+\beta^{\lambda}_{7,1}N+\beta^{\lambda}_{7,2}N^2+\beta^{\lambda}_{7,3}N^3+\beta^{\lambda}_{7,4}N^4+\beta^{\lambda}_{7,5}N^5+\beta^{\lambda}_{7,6}N^6+\beta^{\lambda}_{7,7}N^7.
\label{lamb4a}
\ee
The coefficients of the highest two powers of $N$ may be extracted from the large-$N$ results in Ref.~\cite{bgk,jag1} as
\begin{align}
\beta^{\lambda}_{7,7} =& \tfrac{1}{7}\Bigl[-\tfrac{1}{46656}\zeta_3^2+\tfrac{13}{3359232}\zeta_3+\tfrac{11}{93312}\zeta_5-\tfrac{1}{8817984}\pi^6
+\tfrac{1}{20155392}\pi^4+\tfrac{125}{53747712}\Bigr],\nn
\beta^{\lambda}_{7,6} =& \tfrac17\Bigl[\tfrac{717083}{201553920}-\tfrac{53}{19683}\zeta_3^3+\tfrac{2641}{52488}\zeta_3^2\nn
&+\tfrac{1}{28343520}\left(635391-640\pi^6-26676\pi^4+2749680\zeta_5\right)\zeta_3\nn
&-\tfrac{13}{629856}\left(8\pi^4+2769\right)\zeta_5+\tfrac{8021}{39680928}\pi^6-\tfrac{377}{1166400}\pi^4-\tfrac{31799}{104976}\zeta_7\nn
&-\tfrac{1799}{39366}\zeta_9+\tfrac{1021}{47239200}\pi^8\Bigr].
\label{coeff8}
\end{align}
We have the exact result for $\beta^{\lambda}_6$\cite{schnetz} and so we may use
\be
\delta_6=-\frac{\Acal}{5+X}
\ee 
in our determination of $\Acal$ and $X$. As we have argued earlier in Section~\ref{6phi}, we shall use the highest pair of $\delta$s available for this determination, in this case $\delta_5$ and $\delta_6$. However, as we mentioned in Section~\ref{6phi}, at the eight-loop level it makes little difference to the estimate of $\beta_7^{\lambda}(N)$ which pair of $\delta$s we choose; mostly less than $1\%$ difference in $\beta_7^{\lambda}(N)$ for $N\le 10$, though increasing quite sharply thereafter. In line with this, the values obtained for $\beta_{7,0}^{\lambda}$,  $\beta_{7,1}^{\lambda}$ typically only vary by $1-2\%$ for different choices of pairs of $\delta$s, while the value for $\beta_{7,2}^{\lambda}$ typically varies by $10-20\%$ for different choices; with much more variation for coefficients of higher powers of $N$.

In Table~\ref{phi8} we list the predictions obtained for the coefficients in Eq.~\eqref{lamb4a} using APAP and inputting none, one or two of the known values in Eq.~\eqref{coeff8}. Then in Table~\ref{phi8a} we similarly list the predictions obtained using AAPAP and inputting none, one or two of the known values in Eq.~\eqref{coeff8}.  
We note the astonishing uniformity in the predictions for the first two or even three coefficients; this is similar to the somewhat limited effects of changing the choice of pairs of $\delta$ which we have just observed.

It seemed best to present the full set of results and let the reader make their own choice of the most reliable if desired. But we shall also present our own choice. Based on our observations of Tables~\ref{phi4}, \ref{phi5}, \ref{phi6} and \ref{phi7}, it seems clear that the AAPAP result for the first coefficient is always more accurate. For the second coefficient, AAPAP seems to do better at even loops and APAP at odd loops at lower orders, so we again pick AAPAP; but anyway, the eight-loop results for the second coefficient are very similar for both APAP and AAPAP. The number of inputs also seems to make hardly any difference to these first two coefficients at eight loops. For the third coefficient, AAPAP largely appears to do better at lower orders, especially at six loops which is arguably the best guide to eight loops; and therefore we once again choose AAPAP. The two-input AAPAP prediction is slightly better than the others at six loops, so we pick this one at eight loops too (the eight-loop AAPAP predictions for the third coefficient depends only slightly on the number of input in any case). For the fourth coefficient there is a lot more variation in the lower-loop results, and it is difficult to see a definite pattern; nevertheless, AAPAP with one input does best at five, six and seven loops, so we shall choose this for our eight-loop prediction too (once again, there is anyway little dependence on the number of inputs at eight loops in the AAPAP case). Beyond this level, the accuracies obtained at lower loops are very poor and our eight-loop ``predictions'' vary wildly, so we do not feel there is much point in making a selection. In summary, our most confident results are
\be
\beta^{\lambda}_{7,7} =-2.157\times10^6,\quad \beta^{\lambda}_{7,6} =-1.152\times10^6,\quad \beta^{\lambda}_{7,5} =-204800,\quad 
\beta^{\lambda}_{7,4} =-14430.
\label{8res}
\ee
The first two of these are affected hardly at all by our choice of a pair of $\delta$, and the third very little; and in any case we can argue quite strongly from our lower-order experience that we have made the best choice of $\delta$.
\begin{table}
\begin{center}
\begin{tabular}{|c|c|c|c|}
\hline
&APAP (no input) &APAP($\beta^{\lambda}_{7,7}$)&APAP($\beta^{\lambda}_{7,7}, \beta^{\lambda}_{7,6})$\\
\hline
$\beta^{\lambda}_{7,0}$&$-2.167\times10^6$&$-2.167\times10^6$&$-2.167\times10^6$\\
$\beta^{\lambda}_{7,1}$&$-1.145\times10^6$&$-1.152\times10^6$&$-1.153\times10^6$\\
$\beta^{\lambda}_{7,2}$&$-220800$&$-204800$&$-203800$\\
$\beta^{\lambda}_{7,3}$&$-406.0$&$-13530$&$-14130$\\
$\beta^{\lambda}_{7,4}$&$-5677$&$-486.1$&$-325.2$\\
$\beta^{\lambda}_{7,5}$&$1086$&$16.43$&$-3.644$\\
$\beta^{\lambda}_{7,6}$&$-111.3$&$-0.9392$&I/P\\
$\beta^{\lambda}_{7,7}$&$4.503$&I/P&I/P\\

\hline
\end{tabular}
\end{center}
\caption{\label{phi8}  AAPAP predictions for  eight-loop scalar $\beta$-function coefficients as defined in Eq.~\eqref{lamb4a}, for various inputs}
\end{table}

\begin{table}
\begin{center}
\begin{tabular}{|c|c|c|c|}
\hline
&AAPAP (no input)&AAPAP($\beta^{\lambda}_{7,7}$)&AAPAP($\beta^{\lambda}_{7,7}, \beta^{\lambda}_{7,6})$\\
\hline
$\beta^{\lambda}_{7,0}$&$-2.157\times10^6$&$-2.157\times10^6$&$-2.157\times10^6$\\
$\beta^{\lambda}_{7,1}$&$-1.152\times10^6$&$-1.152\times10^6$&$-1.152\times10^6$\\
$\beta^{\lambda}_{7,2}$&$-204900$&$-204500$&$-204800$\\
$\beta^{\lambda}_{7,3}$&$-14210$&$-14430$&$-14270$\\
$\beta^{\lambda}_{7,4}$&$-360.3$&$-293.2$&$-335.2$\\
$\beta^{\lambda}_{7,5}$&$3.500$&$-7.414$&$-2.270$\\
$\beta^{\lambda}_{7,6}$&$-0.6437$&$0.2413$&I/P\\
$\beta^{\lambda}_{7,7}$&$0.02816$&I/P&I/P\\
\hline
\end{tabular}
\end{center}
\caption{\label{phi8a}  APAP predictions for eight-loop scalar $\beta$-function coefficients, for various inputs (continued)}
\end{table}

We now turn to the task of making predictions for the full eight-loop $\beta$-function $\beta_7^{\lambda}(N)$.  We see in Tables~\ref{phi4acca} and  \ref{phi6acca}  that at four and six loops APAP is better than AAPAP for $N\le L$ (where $L$ is the loop order), while AAPAP is better than APAP for $N >L$, at least for moderate values of $N$. Since we have also argued that the behaviour alternates between odd and even, it seems that the four and six-loop behaviour is our best guide, and accordingly we should use APAP for $N\le8$ and AAPAP for $N>8$.

%We suggested in Section~\ref{6phi} that APAP was more accurate than AAPAP for four, six and seven loops for low values, say $N\le 8$. We also %argued that the even-loop behaviour might be the best guide to eight loops. At larger values of $N$ we suggested that it was sensible to evaluate 
%$\beta^{\lambda}(N)$ using a range $0\le n\le N$, and then we observed that it was less clear that APAP was preferable to AAPAP.

Now let us examine the actual eight-loop data for the full eight-loop $\beta$-function $\beta_7(N)$. Luckily, the major feature will be that there is very little difference between APAP and AAPAP at this loop order.  In Table~\ref{phi8fullb} we give the predictions for $\beta_7^{\lambda}$, using APAP and then AAPAP. We see that the corresponding results for APAP and AAPAP are almost identical up to $N=7$ except for slight variations in the values for $N=2$, and furthermore show little difference between APAP and AAPAP for $N>7$. As explained above, for our best-guess predictions we use the APAP predictions from Table~\ref{phi8fullb} for $[0,8]$ and then select the AAPAP results from Table~\ref{phi8fullb} for $N > 8$, though in any case the APAP/AAPAP difference is likely to be considerably less than the overall error. For completeness, we show the full set of predictions in Table~\ref{phi8fullc}. Again, the first few of these (up to $N\sim 10$) are hardly affected by our choice of pairs of $\delta$, and in any case we would argue that our choice is the best available.

Finally, as we have done in previous cases, we experimented with a negative range. Here we found that APAP gave results with differences from those for a positive range comparable to the differences between APAP and AAPAP, or between the no-input and one-input or two-input cases. Once again, however, we have decided not to incorporate these results, in this case since experience from lower ordersin the $\lambda \phi^4$ case is not at all encouraging.
%As at lower loops, one might surmise that this increased uniformity is due to the fact that in the $\lambda\phi^4$ case we have done a fit to positive values of $N$ and now we are testing positive values of $N$; and indeed if we return to the QCD case, we find that the predictions are in fact very uniform over a wide range of negative values of $N$. 

\begin{table}
\begin{center}
\begin{tabular}{|c|c|c|c|}
\hline
$N$&APAP&AAPAP\\
\hline
0&$-2.167\times10^6$&$-2.157\times10^6$\\
1&$-3.538\times10^6$&$-3.527\times10^6$\\
2&$-5.406\times10^6$&$-5.398\times10^6$\\
3&$-7.867\times10^6$&$-7.867\times10^6$\\
4&$-1.103\times10^7$&$-1.104\times10^7$\\
5&$-1.500\times10^7$&$-1.503\times10^7$\\
6&$-1.992\times10^7$&$-1.997\times10^7$\\
7&$-2.591\times10^7$&$-2.599\times10^7$\\
8&$-3.311\times10^7$&$-3.321\times10^7$\\
9&$-4.169\times10^7$&$-4.181\times10^7$\\
10&$-5.181\times10^7$&$-5.195\times10^7$\\
11&$-6.365\times10^7$&$-6.379\times10^7$\\
12&$-7.739\times10^7$&$-7.753\times10^7$\\
13&$-9.326\times10^7$&$-9.336\times10^7$\\
14&$-1.115\times10^8$&$-1.115\times10^8$\\
15&$-1.322\times10^8$&$-1.322\times10^8$\\
16&$-1.558\times10^8$&$-1.556\times10^8$\\
17&$-1.824\times10^8$&$-1.820\times10^8$\\
18&$-2.125\times10^8$&$-2.117\times10^8$\\
19&$-2.461\times10^8$&$-2.449\times10^8$\\
\hline
\end{tabular}
\end{center}
\caption{\label{phi8fullb}Predictions for  full eight-loop scalar $\beta$-function $\beta_7^{\lambda}(N)$ using APAP and AAPAP and for given values of $N$}
\end{table}

\begin{table}
\begin{center}
\begin{tabular}{|c|c|c|c|}
\hline
$N$&\\
\hline
1&$-3.538\times10^6$\\
2&$-5.406\times10^6$\\
3&$-7.867\times10^6$\\
4&$-1.103\times10^7$\\
5&$-1.500\times10^7$\\
6&$-1.992\times10^7$\\
7&$-2.591\times10^7$\\
%8&$-3.311\times10^7$\\
%9&$-4.169\times10^7$\\
%10&$-5.181\times10^7$\\
%11&$-6.365\times10^7$\\
%12&$-7.739\times10^7$\\
%13&$-9.326\times10^7$\\
%14&$-1.115\times10^8$\\
%15&$-1.322\times10^8$\\
%16&$-1.558\times10^8$\\
%17&$-1.824\times10^8$\\
%18&$-2.125\times10^8$\\
8&$-3.311\times10^7$\\
9&$-4.181\times10^7$\\
10&$-5.195\times10^7$\\
11&$-6.379\times10^7$\\
12&$-7.753\times10^7$\\
13&$-9.336\times10^7$\\
14&$-1.115\times10^8$\\
15&$-1.322\times10^8$\\
16&$-1.556\times10^8$\\
17&$-1.820\times10^8$\\
18&$-2.117\times10^8$\\
\hline
\end{tabular}
\end{center}
\caption{\label{phi8fullc} Selected best-guess predictions for full eight-loop scalar $\beta$-function $\beta_7^{\lambda}(N)$, for given values of $N$}
\end{table}

\section{Conclusions}\label{conc}
We have seen some striking features associated with the Pad\'e predictions. Perhaps the most astonishing is that the predictions for the first three coefficients of the five-loop QCD $\beta$-function (as a power series in $N_F$) which we presented in Ref.~\cite{wap1}, nineteen years before the exact results were obtained in Refs.~\cite{chet1,york1,franz1,york2}, were accurate to $0.2\%$, $1.1\%$ and $6.3\%$ respectively in the one-input case, and $
0.75\%$, $0.82\%$ and $2.4\%$ respectively in the no-input case (see Tables~\ref{qcd5a1} and \ref{qcd5b1}, respectively). These accuracies apply in the ``w/o~QC'' scenario where we discard quartic Casimirs in both the input data and the five-loop coefficients. With the exact results at hand, we could confirm our anticipation that the ``w.~QC'' scenario of retaining quartic Casimirs was much less accurate, though only in the case of the QCD $\beta$-function; there was much less distinction between the two scenarios in the case of the quark mass anomalous dimension. We also performed some ``postdictions'' for the seven-loop $\beta$-function in $O(N)$ $\lambda\phi^4$ theory, achieving using AAPAP in the one-input case (as a typical example) accuracies of $0.04\%$, $0.42\%$ and $1.2\%$ for the first three coefficients respectively in the power series in $N_F$ (see Table~\ref{phi7}). In the scalar case, the total $\beta$-function $\beta^{\lambda}(N_F)$ was also well-predicted and for instance for $N_F=3$ an accuracy of $\sim0.45\%$ was achieved (see Table~\ref{phi7acca}).

In Section~\ref{6qcd} we used the experience gained up to five loops in QCD to make predictions for the six-loop QCD $\beta$-function. One lesson of the five-loop case was that of the three available methods, APAP, AAPAP and WAPAP, none was consistently better than the other two, and WAPAP was disappointing compared with four loops. However, luckily at six loops there was little difference between the results of the three methods, and also between the no-input and one-input cases. One can also detect a somewhat counterintuitive increase in accuracy of the predictions with loop order, so we might expect these convergent predictions to be quite accurate; at least for the first three or four coefficients in the expansion in powers of $N_F$. Our best-guess predictions for the ``w/o~QC'' scenario are listed in Eq.~\eqref{6res}. Despite the poor experience with the ``w.~QC'' scenario at five loops, we felt it best to include the ``w.~QC'' predictions at six loops, both for completeness and in view of the increase in accuracy with loop order which we have noted. Our best-guess predictions for this scenario are presented in Eq.~\eqref{6resa}.

The accuracy of the five-loop predictions for the quark mass anomalous dimensions continued the trend from lower loops in being rather disappointing; but in view of the general increase in accuracy with loop order already noted, we nevertheless presented our six-loop ``w/o~QC'' predictions in Tables~\ref{anom6} and \ref{anom6ctd}, and our six-loop ``w.~QC'' predictions in Table~\ref{anom6a}. Our best-guess ``w/o~QC'' and ``w.~QC'' predictions were given in Eqs.~\eqref{6resAD} and \eqref{6resADa} respectively.

In Section~\ref{6phi} we analysed the accuracy of Pad\'e predictions for the $\beta$-function in $O(N)$ $\lambda\phi^4$ theory. Here, in view of recent advances in perturbative calculations, there is a wealth of accurate data with which to compare our predictions, making it easier to detect trends. Once again we examined the Pad\'e predictions both for the coefficients of powers of $N$ in the polynomial expansion of the $\beta$-function, and also the values $\beta^{\lambda}(N)$ of the $\beta$-function for particular values of $N$. We also noticed that the accuracies at even loops all seemed to be improving with loop order, as did the accuracies at odd loop order, for both the individual coefficients in the $N$ expansion and (somewhat naturally) the full $\beta^{\lambda}(N)$.

In Section~\ref{8phi} we turned to eight-loop predictions in $\lambda\phi^4$ theory. Here a clear convergence was observable in the predictions of the first three or four coefficients in the $N$-expansion, both between APAP and AAPAP and between different numbers of inputs.  As in the case of QCD, this convergence, combined with the increase in accuracy with loop order gives us some confidence in the accuracy of our predictions. As before, we also presented  predictions for $\beta^{\lambda}(N)$ for a range of values of $N$. Our best-guess predictions are given in Eq.~\eqref{8res} and  Table~\ref{phi8fullc}.

In summary we have presented evidence that the Asymptotic Pad\'e Approximation Procedure provides robust and accurate predictions for high-loop renormalisation-group quantities in both QCD and $\lambda\phi^4$ theory. Several questions still remain, however. Possibly the most pressing is to understand why the QCD $\beta$-function (but not the quark mass anomalous dimension) predictions are improved if we drop the new quartic Casimirs both from the inputs to the Pad\'e procedure, and from the exact results with which we compare; and a related question is whether the same would apply to higher Casimirs when they arise. Another interesting issue is to understand why the percentage accuracy appears to increase with loop order, both at even loops and, separately, for odd loops.

\vskip 10pt
\noindent{\bf Data Availability Statement} No data were used in the research described in this article.
\vskip 10pt
\noindent{\bf Acknowledgement} The research was carried out with the support of the
STFC Consolidated Grant ST/X000699/1. The authors thank Y. Schr\"{o}der for useful comments on an earlier
version of the article. For the purpose of open access, the
authors have applied a Creative Commons Attribution (CC-BY) licence to any
Author Accepted Manuscript version arising.

\end{document}